\pdfoutput=1

\documentclass[3p]{elsarticle}

\usepackage{lineno,hyperref}
\usepackage{booktabs}
\usepackage{array} 
\usepackage{paralist} 
\usepackage{verbatim} 
\usepackage{subfig} 
\usepackage{amsfonts}
\usepackage{amsmath}
\usepackage{caption}
\usepackage{placeins}
\usepackage{multirow}
\usepackage{lscape}

\journal{arxiv.org}

\biboptions{authoryear}

\begin{document}

\begin{frontmatter}

\title{GEECORR: A SAS macro for regression models of correlated binary responses and within-cluster correlation using generalized estimating equations}

\author[1]{Tracie L. Shing\corref{cor1}}
\ead{tshing@live.unc.edu}

\author[1]{John S. Preisser}
\ead{jpreisse@bios.unc.edu}

\author[3]{Richard C. Zink}
\ead{richard.zink@lexitas.com}

\cortext[cor1]{Corresponding author}
\address[1]{Department of Biostatistics, Gillings School of Global Public Health,
			University of North Carolina at Chapel Hill,
			Chapel Hill, NC 27599-7420}
\address[3]{Lexitas Pharma Services, Inc., Durham, NC 27701-2102}

\begin{abstract}
A SAS macro, GEECORR, has been developed for the analysis of correlated binary data based on the \citet{prentice_correlated_1988} estimating equations method that extends the \citet{liang_longitudinal_1986} generalized estimating equations (GEE) method to include additional estimating equations for the pairwise correlation between binary variates.  This extension allows for flexible modeling of both the marginal mean and within-cluster correlation as a function of their respective covariate risk factors.  This paper provides an overview of the extended estimating equations method, describes the features and capabilities of the GEECORR macro, and applies the GEECORR macro to three different datasets.  In addition, this paper describes the more detailed fitting algorithm proposed by \citet{prentice_correlated_1988}, of which a variation has been implemented in the GEECORR macro.  We provide a small simulation study to demonstrate the efficiency of the detailed method for estimating correlation parameters.
\end{abstract}


\end{frontmatter}

\section{Introduction}

Binary responses are ubiquitous.  Correlated binary data arise when clusters or repeated measurements of dichotomous responses such as yes/no, presence/absence of disease, or success/failure are collected.  Specifically, units of the same cluster may be more related to each other than to units of different clusters.  Clustering can occur as a result of study designs such as longitudinal measures of the same individual over time, survey designs with cluster sampling, or cluster randomized trials.  Clustering also arises naturally in family studies where measurements are collected for related spouses, parent-child, and sibling pairs or biologically such as teeth within a mouth or eyes of an individual.  There are many scientific research areas such as medicine, epidemiology, and public health in which the correlation between binary responses is of interest in addition to associations of risk factors with the response.  For example, in an oral epidemiological study, characterization of the correlation of dental caries on different teeth found that caries tended to cluster on one side of the mouth which lead to preventive care recommendations \citep{vanobbergen_caries_2007}.  

Generalized estimating equations (GEE) is a semi-parametric method of estimation that is commonly used to fit population-averaged models for correlated outcomes while accounting for within-cluster correlation \citep{liang_longitudinal_1986}. GEE does not require any assumptions about the joint distribution of the response vector beyond the marginal means, which makes it a flexible method that can be used for a variety of outcomes types.  GEE treats correlations as nuisance parameters while inference for marginal mean parameters are robust to misspecification of the assumed or ``working” correlation matrix. However, even when the degree and pattern of clustering are not of primary interest, more accurate modeling of the correlation pattern can result in efficiency gains. The GEE approach is readily available in most statistical software including PROC GENMOD in SAS/STAT software and R packages \textbf{gee} \citep{carey_gee} and \textbf{geepack} \citep{halekoh_geepack}.  These aforementioned statistical software allow for either the choice of a working correlation structure or a user-defined correlation matrix of known values.

When correlation is of interest, fewer methods and even fewer software packages are available for data analysis.  Generalized linear mixed model (GLMM) approaches for binary outcomes are capable of describing between-cluster heterogeneity using random effects and can be implemented using PROC GLIMMIX or PROC NLMIXED in SAS/STAT software or the R package \textbf{glmm} \citep{Knudson_glmm}. However, GLMMs, which have subject-specific interpretations, may not provide suitable inference when the
population-averaged associations of risk factors with the response and marginal model characterizations of the within-cluster association are of interest \citep{preisser_sample_2003}. 

Second-order GEE (GEE2) specifies joint estimating equations to efficiently estimate marginal mean and pairwise association parameters \citep{liang_zeger_qaqish_GEE2_1992}. GEE2 can also be implemented in R using \textbf{geepack} as described in \citet{yan_geepack} and  \citet{yan_fine_geepack}. However, GEE2 does not provide robust inference of the marginal mean model when the within-cluster correlation is misspecified and is computationally intensive, or even prohibitive, for modeling data with large cluster sizes.

There are several estimating equations approaches that maintain the robustness of GEE for inference regarding the marginal mean model while sacrificing some degree of efficiency relative to GEE2.  These approaches uncouple the estimating equations for marginal mean and within-cluster correlations and use a working independence correlation matrix for the association model parameters to speed up computation and more easily handle large cluster sizes. One such method, alternating logistic regressions (ALR), incorporates a model for the within-cluster pairwise odds ratios to describe the association between pairs of responses \citep{carey_alr_1993}.  ALR methods are available in PROC GENMOD in SAS/STAT software and the \textbf{ORTH} package in R \citep{by_orth_2014}.  Other approaches for modeling within-cluster association between outcomes pairs use correlations which are widely reported and used, for example, in power calculations for cluster randomized trials \citep{Teerenstra_2010}.  In particular, the estimating equations method of \citet{prentice_correlated_1988} is widely applicable to modeling marginal means and within-cluster correlations for binary data.  Estimates of asymptotic variances of both marginal mean and correlation parameter estimates are produced facilitating inference.

Therefore, this research article describes a SAS macro GEECORR to implement the \citet{prentice_correlated_1988} method, referred herein as extended GEE, which to our knowledge is not presently available in any statistical software.  The extended GEE method specifies a pair of generalized linear models for flexible modeling of the marginal mean and within-cluster correlation structure.  It estimates parameters with population-averaged interpretations and is computationally feasible for large clusters.  We provide an overview of  Prentice's extended GEE in Section \ref{section:methods}.  In Section \ref{section:macrodetails} we provide additional details of the macro and in Section \ref{section:examples} we demonstrate the application of the GEECORR macro to three data sets.  Finally, we incorporate a more detailed fitting algorithm also described in \citet{prentice_correlated_1988} that yields additional efficiency for variance estimates of the correlation parameters and provide a small simulation study in Section \ref{section:simulation} to investigate the finite sample properties of the two implementations of the extended GEE.

\section{Methods}\label{section:methods}
\subsection{Set-up}
Let $Y_i=(Y_{i1},...,Y_{in_{i}})'$ be the vector of $n_i$ binary responses for the \textit{i}-th cluster and $i=1,...,K$ where there are
$K$ clusters.  Each $Y_{ij}$ is a Bernoulli random variable where $\text{E}[Y_{ij}] = \mu_{ij}$ is the response probability of interest
and $\text{var}[Y_{ij}] = \mu_{ij}(1-\mu_{ij})$.  As such, $\text{E}[Y_{i}] = \mu_i = (\mu_{i1},...,\mu_{in_{i}})'$ is the vector of 
population marginal means for the \textit{i}-th cluster.

Let $X_{ij}$ be a $p \times 1$ vector of covariates for the \textit{j}-th observation in the \textit{i}-th cluster.
Thus, $\mathbf{X}_{i} = (X_{i1},...,X_{in_{i}})$, is a $p \times n_i$ matrix.
In the generalized linear model formulation, $g_{1}(\mu_{ij})=X'_{ij}\boldsymbol{\beta}$ 
where $g_{1}(\cdot)$ is the link function and $\boldsymbol \beta$ is a $p \times 1$ regression parameter vector associated with the marginal mean model.

For observations from the same unit cluster, the correlation between any pair of binary responses is often of interest.
The sample correlation between the \textit{j}-th and \textit{k}-th responses in the \textit{i}-th cluster ($j\ne k$) is
$$
R_{ijk} = \frac{(Y_{ij}-\mu_{ij})(Y_{ik}-\mu_{ik})}{\sqrt{\mu_{ij}(1-\mu_{ij})\mu_{ik}(1-\mu_{ik})}}
$$
which is a random variable with $\text{E}[R_{ijk}] = \rho_{ijk}$, or the population correlation of interest, and
$$
\text{var}[R_{ijk}] = 1 + (1 - 2\mu_{ij})(1 - 2\mu_{ik})(\mu_{ij}(1-\mu_{ij})\mu_{ik}(1-\mu_{ik}))^{-1/2}\rho_{ijk}-\rho_{ijk}^2
$$
For the $m_i = n_i (n_i-1)/2$ pairs in the \textit{i}-th cluster, $R_i=(R_{i12}, R_{i13},...,R_{i(n_{i}-1)n_i})'$ is a $m_i \times 1$ 
vector of correlations.  Additionally,  $\text{E}[R_i] = \rho_i = (\rho_{i12},\rho_{i13},...,\rho_{i(n_{i}-1)n_i})'$ is the vector of 
population correlations for the \textit{i}-th cluster.

Let $Z_{ijk}$ be a $q \times 1$ vector of covariates for the $(j,k)$-th pair where $j<k$ in the \textit{i}-th cluster.  
Thus, $\mathbf{Z}_{i} = (Z_{i12},Z_{i13},...,Z_{i(n_{i}-1)n_i})$, is a $q \times m_i$ matrix.
In the generalized linear model formulation, $g_{2}(\rho_{ijk})=Z'_{ijk}\boldsymbol{\alpha}$ 
where $g_{2}(\cdot)$ is the link function and $\boldsymbol \alpha$ is a $q \times 1$ regression parameter vector associated with the marginal association model.

\subsection{Estimating Equations}

For the marginal mean model, a GEE estimator for $\boldsymbol \beta$ is the solution to the estimating equation
\begin{equation}\label{BetaScore}
	U_\beta = \sum_{i=1}^{K} D_i' V_{i}^{-1} (Y_i - \mu_i) = \mathbf{0}
\end{equation}
where 
$D_i = \frac{\partial \mu_i}{\partial \boldsymbol\beta}$, 
$V_i = A_i C_i(\boldsymbol\alpha) A_i$,  
$C_i(\boldsymbol\alpha)$ is a working correlation matrix for $Y_i$ with parameter vector $\boldsymbol\alpha$, and 
$A_i = \text{diag}(\sqrt{\text{var}(Y_{i1})}, \sqrt{\text{var}(Y_{i2})},..., \sqrt{\text{var}(Y_{in_{i}})})$.
Method of moments estimators for $\boldsymbol \alpha$ are available for various correlation structures.

The extended GEE instead adds that an estimator for $\boldsymbol \alpha$ is the solution to the estimating equations
\begin{equation}\label{AlphaScore}
	U_\alpha = \sum_{i=1}^{K} E_i' W_{i}^{-1} (R_i - \rho_i) = \mathbf{0}
\end{equation}
where $E_i = \frac{\partial \rho_i}{\partial \boldsymbol\alpha}$ and
$W_i = \text{diag}(\text{var}[R_{i12}], \text{var}[R_{i13}], ..., \text{var}[R_{i(n_i-1)n_i}])$.  Note $W_i$ uses an identity working correlation for the samples correlations in order to avoid estimation of additional parameters and limit computational burden by obviating the need to invert $m_i \times m_i$ matrices, $W_i$, for $i=1...,K$.

Estimation of $\boldsymbol \beta$ and $\boldsymbol \alpha$ is achieved by solving the estimating equations \eqref{BetaScore} and \eqref{AlphaScore}, iteratively until convergence.
Combining \eqref{BetaScore} and \eqref{AlphaScore}, the extended GEE quasi-score equations can alternatively be written as
\begin{equation}\label{UsualPrenticeScore}
	U_{\beta,\alpha} = 
	\begin{bmatrix}
		\sum_{i=1}^{K} D_i' V_{i}^{-1} (Y_i - \mu_i)  \\
		 \sum_{i=1}^{K} E_i' W_{i}^{-1} (R_i - \rho_i) 
	\end{bmatrix}
	=
	\sum_{i=1}^{K} 
	\begin{bmatrix}
		D_i' & 0 \\
		0 & E_i'
	\end{bmatrix}
	\begin{bmatrix}
		V_i & 0 \\
		0 & W_i
	\end{bmatrix}^{-1}
	\begin{bmatrix}
		Y_i - \mu_i \\
		R_i - \rho_i
	\end{bmatrix}
		= \mathbf{0}
\end{equation} 

\subsection{Fitting Algorithm}\label{section:MethodUsualPrentice}

The algorithm step used to solve \eqref{BetaScore} is identical to that of \citet{liang_longitudinal_1986}.  Estimation of $\boldsymbol\beta$ is done with iteratively reweighted least squares by regressing $(D_i\boldsymbol\beta + (Y_i - \mu_i))$ on $D_i$, with weight $V_i^{-1}$. 
After the \textit{s}-th iteration, the estimate of $\boldsymbol\beta$ is updated as
\begin{equation}\label{BetaAlgorithm}
	\hat{\boldsymbol\beta}_{s+1} =  \hat{\boldsymbol\beta}_s + \left(\sum_{i=1}^{K} D_i' V_{i}^{-1} D_i\right)^{-1} \sum_{i=1}^{K} D_i' V_{i}^{-1} (Y_i-\mu_i)
\end{equation}

Similarly, the algorithm step used to solve \eqref{AlphaScore} and obtain the estimate of $\boldsymbol\alpha$ after the \textit{s}-th iteration is updated as
\begin{equation}\label{AlphaAlgorithm}
	\hat{\boldsymbol\alpha}_{s+1} =  \hat{\boldsymbol\alpha}_s + \left(\sum_{i=1}^{K} E_i' W_{i}^{-1} E_i\right)^{-1} \sum_{i=1}^{K} E_i' W_{i}^{-1} (R_i-\rho_i)
\end{equation}

Estimation of $\boldsymbol\theta=(\boldsymbol\beta, \boldsymbol\alpha)'$ alternates between updating $ \boldsymbol\beta$ by \eqref{BetaAlgorithm} and $\boldsymbol\alpha$ by \eqref{AlphaAlgorithm}.  Note that $V_i$ in \eqref{BetaAlgorithm} is based on $\rho_i$ computed from the estimate of $\boldsymbol\alpha$ by \eqref{AlphaAlgorithm}.

\subsection{More Detailed Iterative Procedure}\label{section:MethodDetailedPrentice}

\citet{prentice_correlated_1988} also suggested that a more detailed fitting algorithm could be based on the joint asymptotic distribution of $K^{1/2}(\hat{\boldsymbol\beta} - \boldsymbol\beta), K^{1/2}(\hat{\boldsymbol\alpha}-\boldsymbol\alpha)$.   Suppose that the estimating equations for $\boldsymbol \beta$ and $\boldsymbol \alpha$ are not orthogonal such that the algorithm accounts for covariance between $U_\beta$ and $U_\alpha$.  An iteratively reweighted least squares formulation of Fisher's scoring is used to solve the resulting combined estimating equations for $\boldsymbol\theta$.  The algorithm step is updated as
\begin{equation}\label{DetailedAlgorithm}
	\begin{bmatrix}
	\hat{\boldsymbol\beta}_{s+1}\\
	\hat{\boldsymbol\alpha}_{s+1}
	\end{bmatrix}
	=
	\begin{bmatrix}
	\hat{\boldsymbol\beta}_{s}\\
	\hat{\boldsymbol\alpha}_{s}
	\end{bmatrix}
	+
	\begin{bmatrix}
	A & 0\\
	B  & C\\
	\end{bmatrix}
	\begin{bmatrix}
	U_\beta\\
	U_\alpha
	\end{bmatrix}
\end{equation}
where
$$
A = \left(\sum_{i=1}^K D_i' V_i^{-1} D_i \right)^{-1} 
$$
$$ 
B =\left(\sum_{i=1}^K E_i' W_i^{-1} E_i \right)^{-1}
	\left(\sum_{i=1}^K E_i' W_i^{-1} \partial R_i/\partial\beta \right) 
	\left(\sum_{i=1}^K D_i' V_i^{-1} D_i\right)^{-1}
$$
$$
C =  \left(\sum_{i=1}^K E_i' W_i^{-1} E_i \right)^{-1}
$$

Since $\partial R_i / \partial \beta$ is a part of the asymptotic covariance of $(U_\beta, U_\alpha)$, it is reasonable to use the expected information rather than the observed information in $B$, i.e. $\text{E}(\partial R_i / \partial \beta)$.  This will be referred to as the detailed GEE method.  Details of the asymptotic joint distribution derivation and subsequent variance estimators are given in \ref{appendix:jointdistderivation}. In the detailed GEE method, the variance estimators, particularly for $\boldsymbol\alpha$, are slightly different from those applied in the extended GEE method due to the covariance of the estimating equations.  Note that if $U_\beta$ and $U_\alpha$ are orthogonal to one another, then $B = 0$ and \eqref{DetailedAlgorithm} reduces to the extended GEE fitting algorithm in equations \eqref{BetaAlgorithm} and \eqref{AlphaAlgorithm}.

\section{Macro Details}\label{section:macrodetails}
The GEECORR SAS macro was developed to analyze correlated binary data based upon the extended GEE fitting algorithm described in section \ref{section:MethodUsualPrentice}.  However, the macro also provides an option to implement the detailed GEE fitting algorithm described in section \ref{section:MethodDetailedPrentice}.  Table~\ref{table:GEECORROptions} provides a summary of the required and optional inputs of the GEECORR SAS macro.

The macro provides parameter estimates ($\boldsymbol\theta=(\boldsymbol\beta, \boldsymbol\alpha)'$) for link-specified functions of the mean and pairwise within-cluster correlation models in terms of covariates (\textbf{X}) and correlation covariates (\textbf{Z}). 
It has been shown that the uncorrected sandwich variance estimator (BC0) underestimates the variance when the number of clusters is small (e.g., $<$40 or 50 for marginal mean parameters) or moderately small (e.g., $<$80 or 100 for correlation parameters), therefore BC0 and bias-corrected (BC2) standard errors are provided for $\boldsymbol\beta$ and $\boldsymbol\alpha$ estimates \citep{mancl_covariance_2001, preisser_lu_qaqish_2008}.

\subsection{Data Inputs}
Three data set inputs are required to use the macro and preprocessing of typical data sets must be performed.
The first data set input contains a unique cluster identifier, the outcome (y), and marginal mean covariates (\textbf{X}) for all observations.  This is a typical data file format for the analysis of clustered data, with multiple records per cluster.  Note that the outcome must be coded as a 0/1 binary variable, where 1 is the indicator for a positive response.  All other marginal mean covariates should be coded as numeric.  Therefore, recoding of any nominal variables using dummy coding or class indicators must be performed by the user.  If an intercept is to be included in the mean model, a variable equal to 1 for all records must be added to the data set.

The second data set input contains only the marginal correlation covariates (\textbf{Z}), which are explanatory variables related to the cluster pairs.  There is one record for each unique within-cluster pair, resulting in multiple records per cluster.  For the \textit{i}-th cluster of size $n_i$, there are $m_i = n_i (n_i-1)/2$ cluster pairs, resulting in $m_i$ records for the i-th cluster.  For a total of \textit{K} clusters, the entire correlation covariate data set will consist of $\sum_{i=1}^K m_i$ records.  While no identifiers are required for the cluster pair, the correlation covariate data set \textit{must} first be sorted by cluster, to correspond with the outcome and marginal mean covariates data set, then sorted sequentially by the cluster unit pairs such that the first observation corresponds to the pair of the first and second unit. Clusters of size 1 should not have rows in the correlation covariates data set.  If the observations in the correlation covariates data set are not sorted in this manner, it is possible to get incorrect results.  Sample code to create a marginal correlation covariates data set is included in \ref{appendix:dataprocessingcode}.

Finally, the third data set input is a cluster weight data set containing frequencies for each cluster.  If a cluster is representative of other clusters in the data set, then the weight variable represents how many times the cluster is replicated in the sample.  If there is no weighting, that is all clusters are unique, then the cluster weight data set should contain only the cluster identifier and a weight variable equal to 1 for all clusters. 
 
 \subsection{Link Functions for Marginal Mean and Pairwise Correlation Models} 
Options to specify identity, log, and logit link functions for the marginal mean and pairwise correlation models, as well as the Fisher's Z transformation link for the correlation model, are available in the GEECORR macro.  This allows for the direct estimation of marginal mean parameters with different interpretations and expands the scope of correlation models.  

The most commonly applied mean link function for binary data is the canonical logit link due to exponential family properties.  Mean parameters estimated with a logit link are interpreted on the log odds scale and the response probability of interest can be obtained by applying the inverse logit function to the estimates.  Additionally, model-predicted odds ratios for covariate comparisons can be obtained by exponentiating the mean parameter estimates.

When interpretations of explanatory variables through risk ratios are of interest, the log link models covariate effects on the log probability of disease scale.  Thus, model-predicted risk ratios can be obtained by exponentiating the mean parameter estimates.  If the primary interest is the risk difference, then using an identity link directly models covariate effects as differences in the response probability of disease.

Unlike the logit link, the identity and log links for the mean do not necessarily bound the probabilities between 0 and 1 which may lead to convergence issues.  Therefore, during each iteration of the fitting algorithm, the macro checks whether any marginal mean is estimated to be outside of the range of 0 and 1.  If any marginal mean is outside of the range, the macro will exit with an error message.  If an identity link for the mean is specified and there are negative marginal means, a log or logit link is recommended.  If a log link for the mean is specified and if marginal means are greater than 1, then the logit link is recommended.

Many pairwise associations can be modeled as a linear function of correlation covariates through an identity link function.  Sometimes a Fisher's Z transformation link may be applied to bound the correlation parameters between -1 and 1.  The need for a log or logit function may occur when an autoregressive correlation structure, such as $\rho_{ijk} = \alpha^{|j-k|}$, is needed. Using a log link, this can be modeled as $log(\rho_{ijk}) = |j - k|\alpha^{\ast}$ where $\alpha^{\ast}=log(\alpha)$.  Exponentiating the estimate of $\alpha^{\ast}$ yields the autoregressive (AR) correlation estimate $\alpha$. Note that when applying a log link to any pairwise correlation model or a logit link to a model where the pairwise correlation is less than 0.5, parameter estimates for $\boldsymbol\alpha$ will likely be negative.  Therefore, negative initial values for $\boldsymbol\alpha$ should be specified using the STARTALPHA option.  
   
 \subsection{Range Violations and Parameter Shrinking}
There are natural bounds for the correlation between two binary random variables from the same cluster called Fr\`echet bounds 
 \citep{prentice_correlated_1988,qaqish_family_2003,chaganty_range_2006,preisser_comparison_2014}.  For the i-th cluster, let $Y_{ij}$ and $Y_{ik}$ be two Bernoulli random variables where $Y_{ij} \sim \text{Bernoulli}(\mu_{ij})$ and $Y_{ik} \sim \text{Bernoulli}(\mu_{ik})$.  By definitions of probability, $ 0 \le P(Y_{ij} = 1) \le 1$ and $ 0 \le P(Y_{ik} = 1) \le 1$.  Additionally, 
$$ 0 \le P(Y_{ij} = 1, Y_{ik} = 1) \le 1$$
Hence,
$$ \text{max}(0, \mu_{ij} + \mu_{ik} -1) \le P(Y_{ij} = 1 , Y_{ik} = 1) \le \text{min}(\mu_{ij}, \mu_{ik}) $$
If we define $\psi_{ij} = \sqrt{\mu_{ij} / (1-\mu_{ij})}$, it follows that the correlation between $Y_{ij}$ and $Y_{ik}$, or $\rho_{ijk}$ will also be bounded by the Fr\`echet bounds,
\begin{equation}\label{FrechetBounds}
	 \text{max}(-\psi_{ij}\psi_{ik}, - \frac{1}{\psi_{ij}\psi_{ik}}) \le \rho_{ijk} \le \text{min}( \frac{\psi_{ij}}{\psi_{ik}}, \frac{\psi_{ik}}{\psi_{ij}})
\end{equation}

During the iterative update step of the GEE fitting algorithm, the GEECORR macro checks whether the current estimated values for $\boldsymbol\beta$ and $\boldsymbol\alpha$ are consistent with the natural Fr\`echet bounds. The PRINTRANGE=YES option prints details for all range violations that occur during any iteration including cluster and observation identifiers, current mean and correlation estimates, and \eqref{FrechetBounds} for the problematic pairs.  When the PRINTRANGE option is turned off, the macro still details any range violation that occurred during the final estimation iteration.  It is always recommended to check for range violations.

Range violations may lead to convergence issues.  To adjust minor violations, parameter shrinking is available when an identity link is applied to the correlation model.  Parameter shrinking can be performed on $\boldsymbol\alpha$ by specifying SHRINK=ALPHA or on $\boldsymbol\theta = (\boldsymbol\beta, \boldsymbol\alpha)'$ by specifying SHRINK=THETA.  For both options, $\boldsymbol\alpha$ is set to 0 if a range violation occurs during the first iteration.  If subsequent violations occur, $\boldsymbol\alpha := 0.95\boldsymbol\alpha$ for $\boldsymbol\alpha$ shrinking and $ \boldsymbol\theta := \boldsymbol\theta - 0.5^{(r+1)}\delta$ for $\boldsymbol\theta$ shrinking, where $r$ is the current number of shrinking modifications and $\delta$ is the previous update to the estimated parameters.  When range violations lead to convergence issues, it is recommended to try shrinking $\boldsymbol\alpha$ first.  If convergence issues still occur, then $\boldsymbol\theta$ shrinking is possible.   A maximum of 20 shrink modifications per iteration is allowed.  If range violations are still present after 20 modifications, then GEECORR macro will exit the call since estimates are considered unreliable.
 
\subsection{Regression Diagnostics and Output Data sets}
The GEECORR macro provides deletion diagnostics for clusters and observations described by \citet{preisser_perin_deletion_2007}.  These diagnostics extend the GEE diagnostics of \citet{preisser_qaqish_deletion_1996} to include correlation model parameters.
Available regression diagnostics include cluster leverage for $\boldsymbol\beta$ and $\boldsymbol\alpha$  ($\textbf{H}_1$ and $\textbf{H}_2$); cluster and observation influence for $\boldsymbol\beta$ (DBETAC and DBETAO) and cluster influence for $\boldsymbol\alpha$ (DALPHAC); and cluster and observation Cook's Distance for overall model fit (DCLS and DOBS), $\boldsymbol\beta$ ($\mbox{DCLS}_\beta$ and $\mbox{DOBS}_\beta$), and $\boldsymbol\alpha$  ($\mbox{DCLS}_\alpha$ and $\mbox{DOBS}_\alpha$).
  
Regression diagnostics can be requested by specifying an output data set in CLSOUT and/or OBSOUT for cluster and observation level diagnostics respectively.  Predicted probabilities based on the estimated marginal mean model can also be obtained by specifying an output data set in PROBOUT.  The predicted probabilities in PROBOUT are computed based on the estimated marginal mean parameters and subsequently applying the appropriate inverse link function.

\subsection{Correlation Selection Criteria}
Four different selection criterion are included in the standard output of the GEECORR macro: the correlation information criterion (CIC)  \citep{hin_wang_cic_2009}, the trace of the empirical covariance matrix (TECM) \citep{westgate_tecm_2014}, the Gaussian Pseudolikelihood (L\textsubscript{G}) \citep{carey_wang_lg_2011}, and the GEE PRESS Criterion (GPC)\citep{inan_press_2019}.  These criteria have shown to have relatively better performance than their original counterparts, the quasi-likelihood criterion (QIC), the Rotnitzky and Jewell (RJ) criterion, and the Shults and Chaganty (SC) criterion.  Correlation selection criteria is an ongoing area of research in GEE methods.  Other criteria can be programmed via additional SAS/IML software modules using the estimated parameters and covariances under the working correlation of interest.

\section{Examples}\label{section:examples}

In this section, we apply the GEECORR macro to three data sets with binary correlated outcomes that have previously been analyzed.

\subsection{Green Tobacco Sickness}\label{section:ExamplesGTS}
A surveillance study of green tobacco sickness (GTS) among Latino migrant farm workers collected repeated measures from a total of 182 workers from 37 work sites (clusters), over the 12-week growing season from June 15 to September 5, 1999.  The response was a daily indicator for GTS, an occupational illness from nicotine poisoning caused by dermal exposure to tobacco plants and skin absorption of nicotine from them.  The number of workers followed from each camp ranged from 2-9 workers and on average, workers contributed 22 (SD=10) work-days at risk.  The maximum work camp cluster size was 208 worker-days while the minimum cluster size was 20 worker-days.  There are a total of 65 days of GTS out of 4049 observations (total days) in the data set.

Information on GTS risk factors such as type of work (topping, priming, barning, other), years worked in tobacco (first year, 2-4 years, and 5 or more years), worked in wet clothes, max temperature recorded during the day (based on current and previous day), and tobacco use (at least once during the week) were collected during the study.  Since agricultural workers from the same work site often share behaviors and working conditions, workers from the same work site are likely to be more similar to each other than workers from different work sites.  Thus, in this example, there are clusters of workers within a work camp and subclusters of repeated measures from the same worker.

\citet{preisser_gts_2003} previously analyzed the data to identify risk factors and estimate the within-camp and within-worker clustering of GTS using ALR.  In this reanalysis of the GTS data, we instead applied the extended and detailed GEE approaches to obtain estimates of the marginal mean model and the pairwise correlations of GTS, performed GEE regression diagnostics, and demonstrated an analogous analysis using a marginal mean model with a log link.

Let $Y_{ijt}$ be an indicator for GTS for the $t$-th time-point of the $j$-th worker in the $i$-th worker camp.  The marginal model for the probability of GTS including all covariates is
\begin{equation}\label{GTSLogitModel}
	\text{logit}[\text{Pr}(Y_{ijt}=1)] = \beta_0 + \beta_1 x_{1ijt} +  \beta_2 x_{2ijt} +  \beta_3 x_{3ijt} +  \beta_4 x_{4ijt} + 
									 \beta_5 x_{5ijt} +  \beta_6 x_{6ijt} +  \beta_7 x_{7ijt}
\end{equation}
where $x_{1ijt}$ is an indicator for priming work type, $x_{2ijt}$ is an indicator for priming/barning work type, and $x_{3ijt}$ is an indicator for other or barning work type, such that topping work type is the reference group; $x_{4ijt}$ is an indicator for less than 5 years of experience, $x_{5ijt}$ is an indicator for working in wet clothes, $x_{6ijt}$ is the max temperature (in Fahrenheit) recorded during the day centered at 92$^{\circ}$F and divided by 10, and $x_{7ijt}$ is an indicator for tobacco use.

The correlation model for pairs of worker-timepoint observations indexed by $(jt, j't')$ is
\begin{equation}\label{GTSCorrModel}
	\rho_{ijtj't'} = \alpha_1 z_{1ijtj't'} + \alpha_2 z_{2ijtj't'}
\end{equation}
where $\rho_{ijtj't'}$ is the pairwise correlation between time-point observations $(Y_{ijt}, Y_{ij't'})$ and 
$z_{1ijtj't'} =1$ if the pair of time-point observations is from the same worker subcluster $(j=j')$ and 0 otherwise;
$z_{2ijtj't'} =1$ if the pair of time-point observations is from different workers $(j\ne j')$ in the same worker camp and 0 otherwise.

Code used to invoke GEECORR is shown in \ref{appendix:GTSLogitCode}.  Parameter estimates, BC0 and BC2 standard errors for the mean model fitted using extended GEE are shown in Table~\ref{table:GTSMeanParm}.  For the mean model with a logit link, the intercept estimate represents the log odds of GTS among the reference group topping workers, with more than 5 years of experience, who do not work in wet clothes, are not tobacco users, and work at an average temperature of 92$^{\circ}$F.  The risk factor parameter estimates represent increments to the log odds of GTS beyond the reference group.  
Applying the inverse logit function to the intercept, the estimated probability of GTS was 0.0051 (95\% CI: 0.0019, 0.0139) in the reference group.  BC2 were slightly larger than BC0 for all parameter estimates.  For a small number of clusters, BC0 tends to underestimate the variance.  Therefore, the larger BC2 standard errors may be appropriate for a moderate number of clusters such as K=37.

Risk factor odds ratio estimates are shown in Table~\ref{table:GTSORRR} with confidence intervals computed based on BC2.  According to the extended GEE analysis, priming workers had 3.54 (95\% CI: 1.68, 7.47) times the odds of GTS compared to topping workers, workers who worked in wet clothes had 2.53 (95\% CI: 1.15, 5.58) times the odds of GTS compared to workers who did not work in wet clothes, and a 10$^{\circ}$F increase in temperature was associated with 1.55 (95\% CI: 1.03, 2.34) times the odds of GTS.  Relative to topping workers, the odds of GTS among priming/barning workers and other or barning workers were not significantly different.  Work experience and tobacco use were also not significantly associated with the odds of GTS.

The estimated within-worker and within-camp correlations using the extended GEE method are shown in Table~\ref{table:GTSCorrParm}.  The within-worker correlation was found to be 0.0261, while the within-camp correlation was found to be 0.0143.  The relative direction of the point estimates agree with the pairwise odds ratios estimated by ALR which found that the odds ratio of GTS between any two days measured for the same worker was higher than the odds ratio of GTS between any two days measured for different workers \citep{preisser_gts_2003}.  The estimates are both less than 0.1 suggesting small worker and camp clustering effects, which could be attributed, in part, to the low prevalence of GTS in the study population.

\subsubsection{Application of the detailed GEE}
The mean and correlation models were also fit using the detailed GEE method described in section \ref{section:MethodDetailedPrentice} and the code shown in \ref{appendix:GTSLogitDetailedCode}.  Parameter estimates, BC0 and BC2 standard errors for the mean model fitted using the detailed GEE were the same as those fitted with the extended GEE method to three decimal places shown in Table~\ref{table:GTSMeanParm}.  Thus, the odds ratio estimates for risk factor comparisons in Table~\ref{table:GTSORRR} were also equivalent after appropriate rounding. The correlation estimates from the extended and detailed GEE methods are also the same as shown in Table~\ref{table:GTSCorrParm}.  BC0 and BC2 standard errors for the within-worker correlation from the detailed GEE method decreased by 23\% and 24\% from the extended GEE method while BC0 and BC2 standard errors for the within-camp correlation decreased by 18\% and 17\%.  The variance estimators in the detailed GEE incorporate the expected information from the score equation for $\boldsymbol\alpha$ with respect to $\boldsymbol\beta$, thereby slightly increasing the efficiency of $\boldsymbol\alpha$ which is advantageous for power calculations.

\subsubsection{Regression diagnostics applied to the logit link for the marginal mean model}
\begin{figure}
	\caption{Cluster-level Cook's Distance for GTS Study}\label{fig:GTSClusterCookD}
	\includegraphics{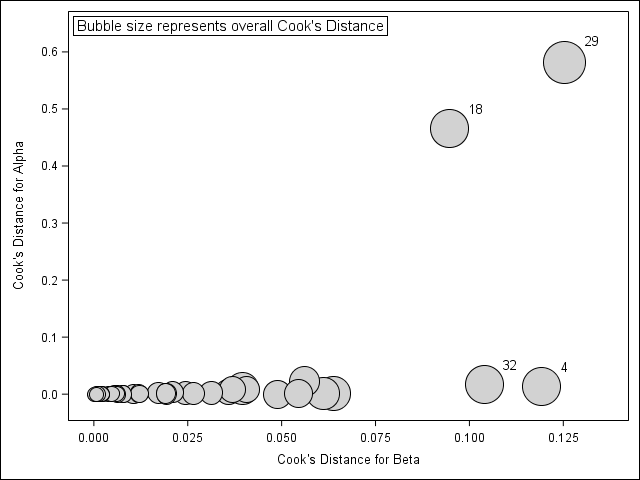}
\end{figure}

Regression diagnostics were performed for both mean and correlation parameters from models \eqref{GTSLogitModel} and \eqref{GTSCorrModel} estimated using the extended GEE.  Figure~\ref{fig:GTSClusterCookD} shows the cluster-level Cook's distance for $\boldsymbol\beta$, $\boldsymbol\alpha$, and overall.  Clusters 4, 18, 29 and 32 had the largest overall Cook's distance.  Clusters 18 and 29 seemed to influence both $\boldsymbol\beta$ and $\boldsymbol\alpha$ while clusters 4 and 32 only appeared to influence $\boldsymbol\beta$.

\begin{figure}
	\caption{Cluster Deletion Diagnostics for GTS Exposure Risk Factors}\label{fig:GTSDBETAC}
	\subfloat[Priming]{\includegraphics[width=0.48\textwidth]{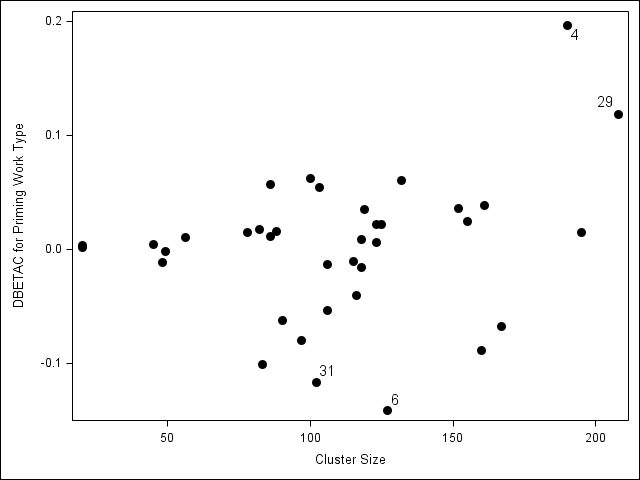}}\hfill
	\subfloat[Priming/barning]{\includegraphics[width=0.48\textwidth]{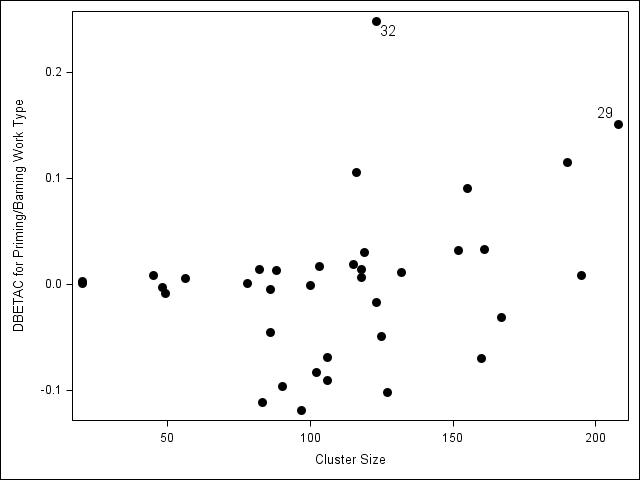}}\\[-2ex]
	\subfloat[Other or barning]{\includegraphics[width=0.48\textwidth]{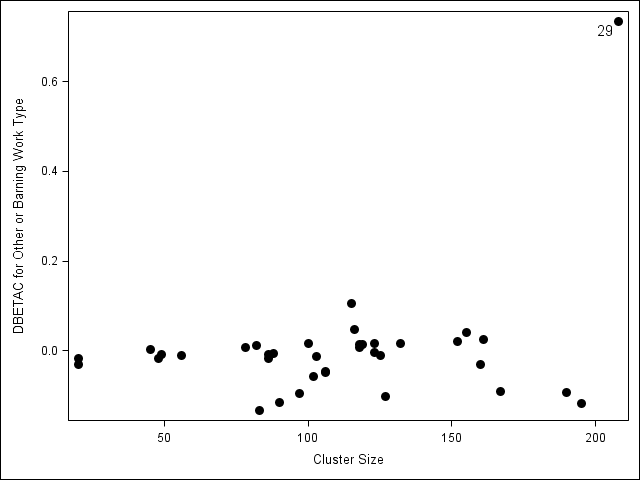}}\hfill
	\subfloat[Less than 5 years of experience]{\includegraphics[width=0.48\textwidth]{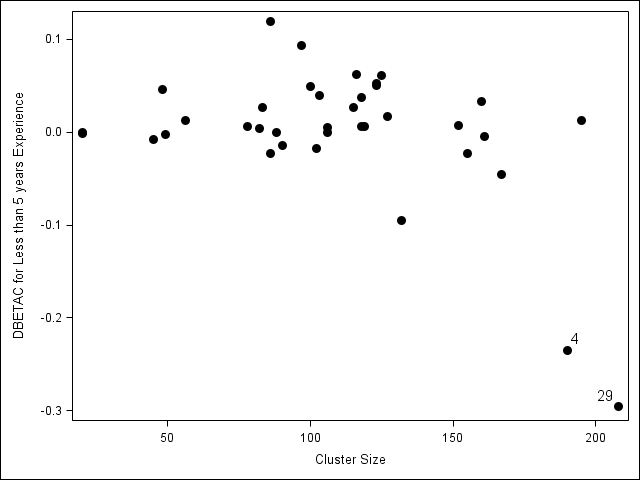}}\\[-2ex]
		\subfloat[Worked in wet clothing]{\includegraphics[width=0.48\textwidth]{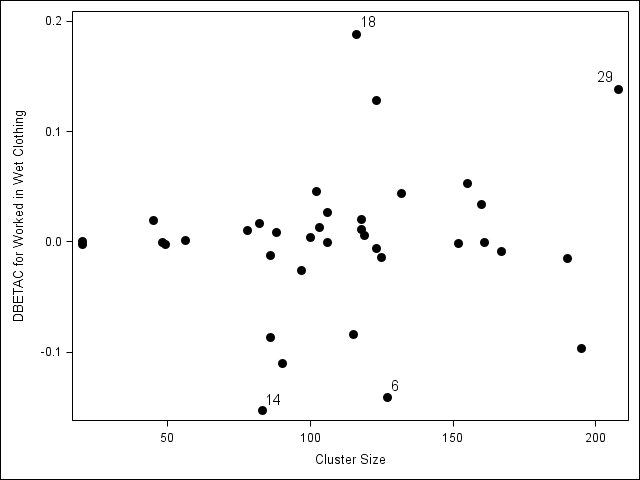}}\hfill
	\subfloat[Temperature (1 unit = 10$^{\circ}$F)]{\includegraphics[width=0.48\textwidth]{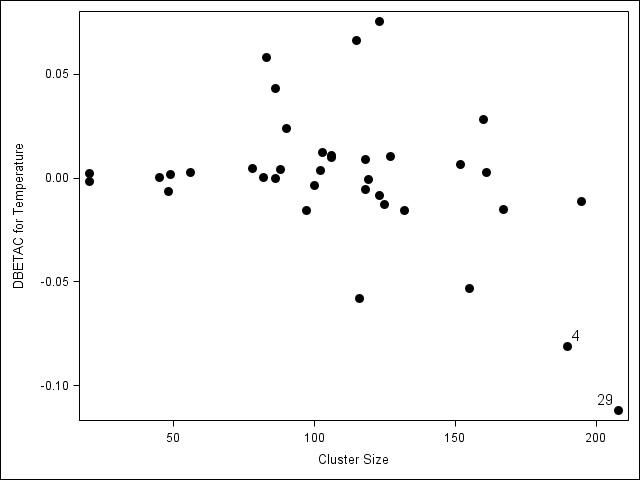}}
\end{figure}

Cluster deletion influences on the GTS risk factor parameter estimates are plotted by cluster size in Figure~\ref{fig:GTSDBETAC}.  Cluster 29 had the largest cluster size and appeared to have a high influence on all risk factor estimates.  Cluster 4 also had a large cluster size and seemed to influence the parameter estimates for the effects of priming over topping, work experience, and temperature.  Clusters 18 and 32 had moderate cluster sizes but had large influences on the parameter estimates for worked in wet clothing and the effect of priming/barning over topping, respectively.  This influence is not surprising considering all days of GTS from cluster 18 occurred among workers who worked in wet clothing and all days of GTS from cluster 32 occurred among priming/barning workers.

\subsubsection{Application of log link for the marginal mean model to estimate risk ratios}
We repeated the analysis by replacing the logit link with the log link for the marginal mean model for the probability of GTS in equation \eqref{GTSLogitModel} in order to directly estimate parameters with interpretations as risk ratios and used the GEECORR macro call shown in \ref{appendix:GTSLogCode}.  
Parameter estimates, BC0 and BC2 standard errors for the mean model with a log link and fitted using extended GEE are also shown in Table~\ref{table:GTSMeanParm}.  The intercept estimate represents the log probability of GTS among the reference group.  The risk factor parameter estimates represent increments to the log probability of GTS beyond the reference group.  Exponentiating the intercept, the estimated probability of GTS was 0.0052 (95\% CI: 0.0019, 0.0142) in the reference group, which is nearly equivalent to the estimated probability from the logit link model.  The risk factors parameter estimates differ slightly from those estimated under a mean model with a logit link.

Risk ratio estimates for the marginal mean model fitted with the log link are shown in Table~\ref{table:GTSORRR}.  According to the extended GEE analysis, priming workers had 3.39 (95\% CI: 1.64, 7.04) times the risk of GTS compared to topping workers, workers who worked in wet clothes had 2.48 (95\% CI: 1.14, 5.40) times the risk of GTS compared to workers who did not work in wet clothes, and a 10$^{\circ}$F increase in temperature was associated with 1.51 (95\% CI: 1.01, 2.26) times the risk of GTS.  Relative to topping workers, the risk of GTS among priming/barning workers and other or barning workers were not significantly different.  Work experience and tobacco use were also not significantly associated with the risk of GTS.

In epidemiological surveillance studies, the interpretation of risk ratios may be more relevant than odds ratios in evaluating risk factors related to disease and on some occasions, the risk ratios may be reported instead of the odds ratios.  However, since GTS is a ``rare outcome'' with the probability of disease less than 1\%, the odds and risk of disease are approximately the same, which is also demonstrated by the similarities between the parameter estimates in Table~\ref{table:GTSMeanParm} and odds ratios and risk ratios in Table~\ref{table:GTSORRR}.
 
The same correlation model in equation \eqref{GTSCorrModel} was used in this re-analysis with a log link marginal mean model.  The estimated correlation parameters were identical to the results from the logit link model after rounding.  Without rounding, the estimated correlation parameters differed at the 4th decimal place.  This is expected due to the different $\boldsymbol\beta$ estimates that are updated in the fitting algorithms.  Similar trends in the parameter estimates occurred when fitting the marginal mean with a log link using detailed GEE compared to extended GEE.

\subsection{CARDIA}

The longitudinal Coronary Artery Risk Development in Young Adults (CARDIA) study is a population-based multicenter cohort study that began collecting the binary outcome for smoking status in 1986.  We re-visited the 15-year CARDIA data assessed by \citet{perin_cardia_2009} and \citet{inan_press_2019} that both used the GEE approach to evaluate changes in cigarette smoking prevalence in four race/sex groups: black females (K=1473), black males (K=1145), white females (K=1299), and white males (K=1160).

In the 15-year data, binary smoking status (yes/no) was assessed at a baseline visit (year 0) and five follow-up visits in 2, 5, 7, 10, and 15 years after the study start.  In this re-analysis, we used extended GEE to directly estimate the correlation parameters and their standard errors rather than treat them as nuisance and posited additional correlation models incorporating unequally spaced time.

Let $Y_{ij}$ be an indicator for current smoking status for the $i$-th subject at the $j$-th timepoint $(j=1,...,6)$.  We used the same marginal model proposed by \citet{inan_press_2019},
\begin{equation}\label{CARDIAMeanModel}
	\text{logit}[\text{Pr}(Y_{ij}=1)] = \beta_0 +\beta_1 x_{1i} + \beta_2 x_{2i} + \beta_3 x_{3i} + \beta_4 x_{4i} + 
	\beta_L year_{ij} + \beta_Q year_{ij}^2 + \beta_C year_{ij}^3
\end{equation}
where $x_{1i}$ = age in years at baseline divided by 10, $x_{2i} = x_{1i}^2$, $x_{3i} = 1$ if highest attained education at baseline is some college without degree and 0 otherwise, $x_{4i} = 1$ if highest attained education at baseline is college degree and 0 otherwise, and $year_{ij}$ is years since the first study exam in 1986 divided by 10.  Note that $x_{3i} = 0$ and $x_{4i} = 0$ denotes the reference education group of no college attended.

Since \citet{perin_cardia_2009} selected an exchangeable working correlation between time points based on exploratory analysis and \citet{inan_press_2019} additionally applied an AR(1) working correlation, we considered the following correlation models that can be implemented as a linear function of explanatory variables related to the cluster pair.

An exchangeable correlation model,
\begin{equation}\label{CardiaCorrExch}
	\rho_{ijk} = \alpha_0 
\end{equation}
a modified AR(1) correlation model using unequal time,
\begin{equation}\label{CardiaCorrAR}
	\rho_{ijk} = \alpha^{|(year_k - year_j)/4|}
\end{equation}
which we model via the log link,
a 5-dependent correlation model using all equally spaced visits,
\begin{equation}\label{CardiaCorrM}
	\rho_{ijk} = \alpha_{k-j}
\end{equation}
and a correlation model using unequal time,
\begin{equation}\label{CardiaCorrUneq}
	\rho_{ijk} = \alpha_{year_k - year_j}
\end{equation}
where $\rho_{ijk}$ is the pairwise correlation between $(Y_{ij}, Y_{ik})$ such that $j<k$.  The correlation structures in equations \eqref{CardiaCorrExch} and \eqref{CardiaCorrAR} consist of one parameter, while \eqref{CardiaCorrM} and \eqref{CardiaCorrUneq} consist of five and eight parameters, respectively.  Unequal time in \eqref{CardiaCorrAR} was scaled by a factor of 4 to minimize potential underflow problems.

The marginal mean model was fit in combination with the four correlation models for each of the four race/sex groups.  All models were fit using extended GEE using the macro calls shown in \ref{appendix:CARDIAExchCode}-\ref{appendix:CARDIAUneqCode}.  Various working correlation selection criteria for each race/sex group are shown in Table~\ref{table:CARDIACorrSelection}.  The Lg, CIC, and TECM selected either the 5-dependent or the unequal time correlation structures for all race/sex groups.  The GPC selected the simpler exchangeable correlation for females and white males, but unequal time for black males. The modified AR(1) correlation was not selected by any of the four criteria.

\begin{figure}[htbp]
	\caption{Exchangeable Correlation for CARDIA Study}\label{fig:CARDIAExchCorrHeatmap}
	\includegraphics{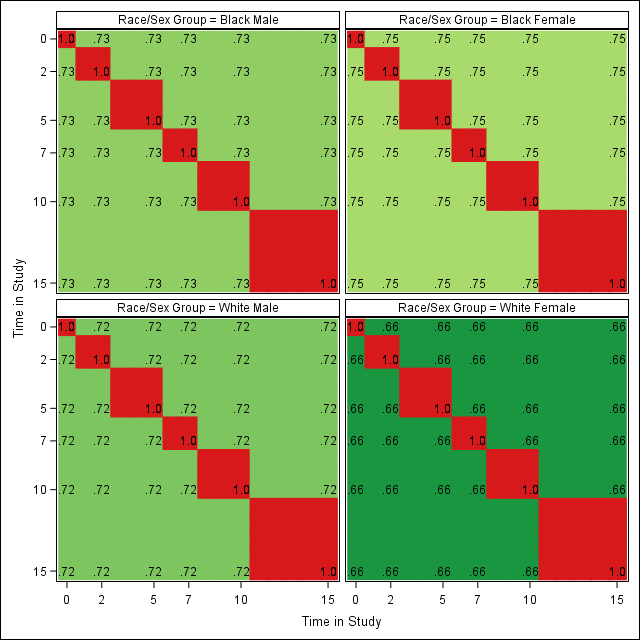}
\end{figure}

\begin{figure}[htbp]
	\caption{AR(1) Correlation for CARDIA Study}\label{fig:CARDIAARCorrHeatmap}
	\includegraphics{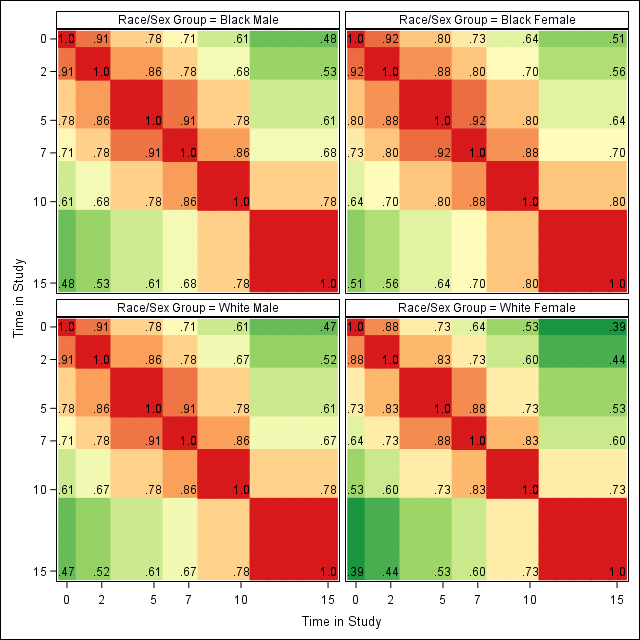}
\end{figure}

\begin{figure}[htbp]
	\caption{5-Dependent Correlation for CARDIA Study}\label{fig:CARDIAMDepenCorrHeatmap}
	\includegraphics{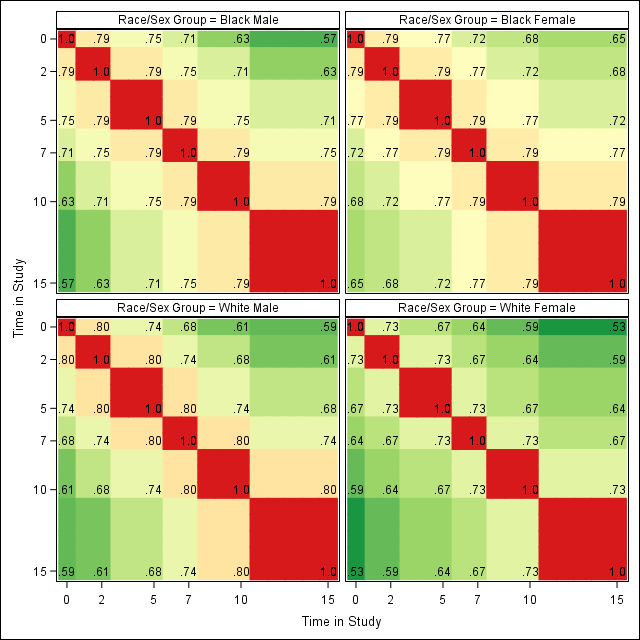}
\end{figure}

\begin{figure}[htbp]
	\caption{Unequal Time Correlation for CARDIA Study}\label{fig:CARDIAUneqCorrHeatmap}
	\includegraphics{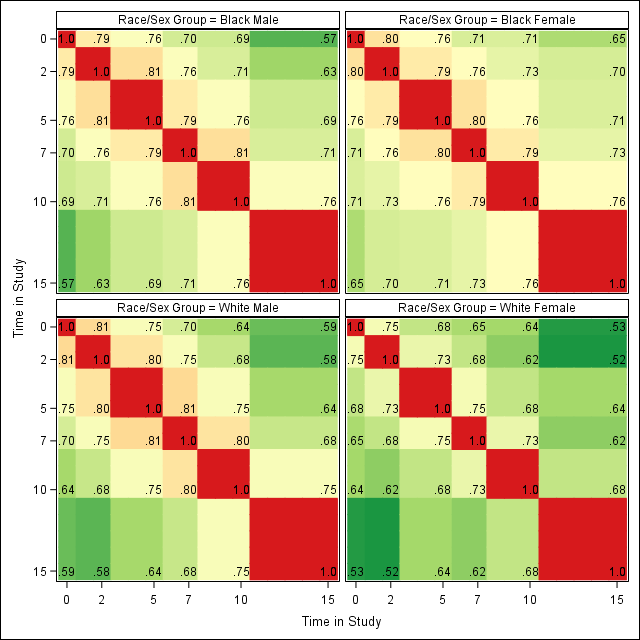}
\end{figure}

The estimated correlation patterns are shown in Figures~\ref{fig:CARDIAExchCorrHeatmap}-\ref{fig:CARDIAUneqCorrHeatmap}.  
All correlation models had the same race/sex group pattern in which white females had the lowest estimated correlations between visits, black females had the highest estimated correlations, and correlations for both male groups were similar to that of black females.  The estimated exchangeable correlation for all race/sex groups ranged from 0.66 to 0.75.  However, Figures~\ref{fig:CARDIAARCorrHeatmap}-\ref{fig:CARDIAUneqCorrHeatmap} demonstrate noticeable visit-based differences in correlation that are not captured in the exchangeable model.  Comparing the correlation patterns for black males, the modified AR(1) correlations ranged from 0.48 for a 15 year difference to 0.91 for a 2 year difference in time.  The spread of correlation estimates from the 5-dependent and unequal time models were narrower than the modified AR(1) correlations, ranging from 0.57 to 0.79 and 0.57 to 0.81, respectively.  Although the modified AR(1) model incorporates time, the selection criteria further suggests that the moderate 5-dependent and linear unequal time correlations may be more appropriate.

Selection between the 5-dependent and linear unequal time correlation structures is nearly equivalent. Using \eqref{CardiaCorrM}, the correlations are the same between visit, regardless of the time.  This is shown by the constant pattern of the diagonals in Figure~\ref{fig:CARDIAMDepenCorrHeatmap}.  In contrast, by incorporating unbalanced time between visits using \eqref{CardiaCorrUneq}, there are noticeable differences between visits with greater lengths of time such as the 3-year correlation (0.81) compared to the 5-year correlation (0.76) as seen on the diagonals of Figure~\ref{fig:CARDIAUneqCorrHeatmap}.  If marginal mean interpretations are of primary interest, the more parsimonious 5-dependent correlation model might be sufficient.  If the correlations are of interest, then there is relevant information in determining the unequal time-based correlations.

Parameter estimates and robust standard errors for the marginal mean for all models are shown in Table~\ref{table:CARDIAMeanEst}.  Although the differences are minor, the mean parameters estimated in conjunction with the 5-dependent and unequal time correlation models generally had smaller standard errors compared to the mean parameters estimated in combination with the exchangeable and modified AR(1) correlation models, which illustrates that selection of the appropriate correlation structure may lead to better efficiency in the mean estimates.  Regardless of the correlation structure, young adults with some college or a college degree significantly decreased the odds of smoking compared to those who did not attend any college in all race/sex groups.  According to the models with a 5-dependent correlation structure, black females who attended some college had 49\% decreased odds of smoking, black males had 61\% decreased odds, white females had 44\% decreased odds, and white males had 63\% decreased odds compared to those who did not attend college in their respective race/sex group.  Meanwhile, black females who obtained a college degree had 83\% decreased odds of smoking, black males had 83\% decreased odds, white females had 84\% decreased odds, and white males had 86\% decreased odds compared to those who did not attend any college in their respective race/sex group.

\subsection{NHANES}

The final example demonstrates an application of the extended GEE to the characterization of chronic periodontitis.  While periodontitis is a common oral health condition defined at the subject (i.e., person) level, the clinical diagnosis of health or disease is made at the individual tooth-site level based on a combination of measures of pocket depth (PD), clinical attachment level (CAL), and bleeding on probing.  For a single patient, a full mouth clinical exam includes the inspection of six sites (mesiobuccal, buccal, distal buccal, mesiolingual, lingual, or distal lingual) on each of 28 teeth (third molars are usually excluded) for a total of 168 tooth sites with possible periodontal disease.  This results in the unique problem of characterizing periodontal disease through both the distribution of tooth-site level probability of disease as well as the natural clustering of tooth sites in the mouth. 

Data from the full-mouth periodontal examination (FMPE) of the 2013-2014 National Health and Nutrition Examination Survey (NHANES) was analyzed.  Although definitions of periodontal disease vary, for an illustration of the analysis of correlated binary data we defined tooth-site level disease as having a probing depth (PD) measurement of 4 mm or greater.

Let $Y_{ij}$ be an indicator for PD greater than or equal to 4mm for the $j$-th numbered tooth site $(j=1,...,n_i), n_i \le 168$ of the $i$-th mouth. Dividing the human dentition into sextants: maxillary right posterior, maxillary anterior, maxillary left posterior, mandibular left posterior, mandibular anterior, and mandibular right posterior and considering that disease may vary between all six tooth sites, we defined the marginal mean model including the main effects for sextants and tooth site.
\begin{align}\label{NHANESMeanModel}
	\text{logit}[\text{Pr}(Y_{ij}=1)] =  \beta_0 & + \beta_1 x_{1ij} + \beta_2 x_{2ij} + \beta_3 x_{3ij} + \beta_4 x_{4ij} 
									 + \beta_5 x_{5ij}  \nonumber \\ 
								 & + \beta_6 x_{6ij} + \beta_7 x_{7ij} + \beta_8 x_{8ij} + \beta_9 x_{9ij} 
									 + \beta_{10} x_{10ij} 
\end{align}
where $x_{1ij}$ is an indicator for a site located on a tooth in the maxillary right posterior,  $x_{2ij}$ is an indicator for a site located on a tooth in the maxillary anterior,  $x_{3ij}$ is an indicator for a site located on a tooth in the maxillary left posterior,  $x_{4ij}$ is an indicator for a site located on a tooth in the mandibular left posterior,  $x_{5ij}$ is an indicator for a site located on a tooth in the mandibular anterior, $x_{6ij}$ is an indicator for a mesiobuccal site, $x_{7ij}$ is an indicator for a buccal site, $x_{8ij}$ is an indicator for a distal buccal site, $x_{9ij}$ is an indicator for a mesiolingual site, and $x_{10ij}$ is an indicator for a lingual site.  The mandibular right posterior and distal lingual tooth site are the reference groups for sextant and tooth site main effects, respectively.  Therefore, $\beta_0$ is the log odds of PD greater than or equal to 4mm among distal lingual sites located in the mandibular right posterior of the mouth.

We considered a correlation model including both spatial factors such as tooth adjacency (teeth that are next to each other), vertical tooth adjacency (teeth that are on top of each other), site adjacency (tooth-sites that are next to each other) and site factors such as sites on the same tooth or sites that share the same interproximal (IP) space between teeth.   The following correlation model was used to describe the disease pattern between tooth sites based on six exhaustive and mutually exclusive categories:
\begin{equation}\label{NHANESCorrModel}
	\rho_{ijk} = \alpha_1 z_{1ijk} + \alpha_2 z_{2ijk} + \alpha_3 z_{3ijk} + \alpha_4 z_{4ijk} + \alpha_5 z_{5ijk} + \alpha_6 z_{6ijk}
\end{equation}
where $\rho_{ijk}$ is the pairwise correlation between $(Y_{ij}, Y_{ik})$, or the \textit{j}-th and \textit{k}-th tooth sites, such that $j<k$ for the i-th subject.  $z_{1ijk}$ is an indicator that the $(j,k)$ pair of sites are adjacent (on the same tooth or on adjacent teeth) and share the same IP space, $z_{2ijk}$ is an indicator that the $(j,k)$ pair of sites are adjacent (on the same tooth) and do not share the same IP space, $z_{3ijk}$ is an indicator that the $(j,k)$ pair of sites are non-adjacent but on the same tooth, $z_{4ijk}$ is an indicator that the $(j,k)$ pair of sites are non-adjacent but on adjacent teeth on the same jaw,  $z_{5ijk}$ is an indicator that the $(j,k)$ pair of sites are on vertically adjacent teeth on the different jaws, and $z_{5ijk}$ is an indicator that the $(j,k)$ pair of sites are non-adjacent teeth.   

Code used to the invoke GEECORR is shown in \ref{appendix:NHANESCode}.  Parameter estimates for the marginal mean and correlation models, fitted with extended GEE, are shown in Table~\ref{table:NHANESResults}.  Among all pairwise correlation types, the correlation between adjacent tooth sites was the highest with estimates of 0.338 (SE=0.028) for sites that shared the same IP space and 0.295 (SE=0.029) for sites that did not share the same IP space.  Site pairs on non-adjacent teeth and site pairs on vertically adjacent teeth had the lowest correlations of 0.148 (SE=0.019) and 0.178 (SE=0.021).  Pairs of non-adjacent sites that were on the same tooth or adjacent teeth were moderately correlated compared to adjacent site pairs and site pairs on non-adjacent teeth.

Odds ratio estimates for comparisons of tooth location and tooth site risk factors are also shown in Table~\ref{table:NHANESResults}. 
Compared to sites on mandibular posterior right teeth, sites located on maxillary anterior teeth (OR=0.60, 95\% CI: 0.53, 0.68)  and mandibular anterior teeth (OR=0.51, 95\% CI: 0.46, 0.57) had significantly decreased odds of disease.  This suggests that there may be increased odds of disease in the back teeth compared to the front teeth.

There were no conclusive differences in the odds of disease between sites on maxillary posterior right or mandibular left posterior teeth and mandibular posterior right teeth.  However, there was some evidence to suggest that the odds of disease is higher for sites on maxillary posterior left teeth compare to mandibular posterior right teeth.  Thus, the odds of disease appeared to be similar in the sextants directly above and across from the mandibular posterior right sextant, but higher in the spatially diagonal maxillary posterior left sextant.

All tooth sites had significantly decreased odds of disease compared to the distal lingual tooth site, which is known to be one of the more difficult tooth sites to reach during brushing.  Sites on the buccal side of the teeth appeared to have decreased odds of disease compared to their lingual counterparts.  Sites located in the IP space, mesio- or distal- sites, appeared to have increased odds of disease compared to the middle tooth sites.

\section{Simulation}\label{section:simulation}

\subsection{Simulation Design}

A simulation experiment was conducted to compare the relative performances of the extended and detailed GEE methods. The simulation study design was motivated by the GTS surveillance study, described in section \ref{section:ExamplesGTS}.  Three simulations were performed by replicating the design $(X_i; Z_i)$ using the original $K=37$ clusters of the GTS study, then doubling and tripling the clusters.

Based on the covariate distribution of the original GTS study, correlated binary data were generated for the GTS marginal mean model in \eqref{GTSLogitModel} and correlation model in \eqref{GTSCorrModel} using the method of \citet{emrich_piedmonte_1991}.  Except for the intercept, $\beta = (-3.9, 1.3, 0.4, -0.7, 0.4, 0.9, 0.4, -0.4)'$ and $\alpha=(0.03, 0.01)'$ were fixed for all simulations to be similar to the analysis in section \ref{section:ExamplesGTS}. The intercept value of -3.9 was chosen to increase the average probability of GTS to 5\% to minimize complete or quasi-complete separation in the simulations.  For each simulation, 1000 random samples were generated and models \eqref{GTSLogitModel} and \eqref{GTSCorrModel} were fitted using both extended and detailed GEE.

Percent relative bias and coverage were used to assess the performances of the fitting algorithms and variance estimators.  Percent relative bias of $\hat{\boldsymbol\beta}$ and $\hat{\boldsymbol\alpha}$ was computed as
$$
(1/N_R) \sum_{r=1}^{N_R} [(\hat{Q}_r - Q_T)/Q_T] \times 100
$$
where $N_R$ is the number of replicates that converged, $\hat{Q}_r$ is $\hat{\boldsymbol\beta}$ or $\hat{\boldsymbol\alpha}$ from the \textit{r}-th simulated replicate and $Q_T$ is the true value of the parameter.
Additionally, percent relative bias of the sandwich and bias-corrected variance estimators was calculated as
$$
\left[\sum_{r=1}^{N_R}\{ \widehat{\text{var}}_s (\hat{Q}_r) \} / N_R - \text{var}_\text{MC}(\hat{Q})\right] \Big/  \text{var}_\text{MC}(\hat{Q}) \times 100
$$
where \textit{s} indexes either BC0 or BC2 and the Monte Carlo simulation variance is
$$
\text{var}_\text{MC}(\hat{Q}) = \sum_{r=1}^{N_R}\{ \hat{Q}_r  - \sum \hat{Q}_r/N_R\}^2 \Big/ (N_R-1)
$$
Finally, coverage was defined as the percent of 95\% confidence intervals for $Q_r$, computed as
$
\hat{Q}_r \pm Z_{0.975} \times \widehat{\text{var}}_s^{1/2}(\hat{Q}_r)
$
that contained $Q_T$.

\subsection{Simulation Results}

Results from the simulation are shown in Table~\ref{table:SimResults}. When $K=37$, 98\% of the replicates converged for both fitting algorithms.  All replicates converged when $K=74$ or 111.

The average marginal mean parameter estimates from 1000 replicates were identical between the extended and detailed GEE methods after rounding resulting in equivalent percent relative biases of the marginal mean parameter estimates. The absolute percent relative biases were less than 5\% for all marginal mean model covariates for 37 clusters, less than 3\% for 74 clusters, and less than 2\% for 111 clusters.  The percent relative bias for both correlation parameters decreased as the number of clusters increased.

Performance of BC0 and BC2 for all marginal mean parameters were comparable between the extended and detailed GEE methods.  Generally, the coverage of both BC0 and BC2 for all marginal mean parameters increased as the number of clusters increased.  In addition, the coverage of BC2 was slightly larger than BC0, regardless of the number of clusters.
With 37 clusters, the BC0 underestimated the Monte Carlo simulation variance of all marginal mean parameters by between 6-17\%.  However, the percent relative bias of BC0 for the marginal mean parameters appeared to trend towards zero as the number of clusters increased, performing the best for 111 clusters with absolute percent relative biases less than 6\%.
Overall, BC2 performed well regardless of the cluster size.

The performance of BC0 and BC2 for the correlation parameters was noticeably different between the extended and detailed GEE methods.  As expected, the coverage of both variance estimators for the correlation parameters increased as the number of clusters increased for both methods. The coverages were slightly lower when fitted with the detailed GEE compared to the extended GEE.  However, variance estimators from the extended GEE method appeared to overestimate the Monte Carlo variance, especially for the within-worker correlation.  Thus, the larger coverages from the extended GEE method may in part be due to this overestimation.

For the extended GEE method, neither BC0 nor BC2 performed well.  BC0 overestimated the Monte Carlo variance of the within-worker correlation ($\alpha_1$) and actually increased as the number of clusters increased.  For the within-camp correlation ($\alpha_2$) the percent relative bias of BC0 appeared to fluctuate about zero and performed the best when there were 74 clusters.  BC2 overestimated the Monte Carlo variance for both the within-worker and within-camp correlations and the percent relative bias also increased as the number of clusters increased.

In contrast, for the detailed GEE method, BC0 and BC2 performed better as the number of clusters increased.  Both variance estimators, underestimated the Monte Carlo simulation variance of the correlation parameters, but the percent relative biases of the variance estimators trended towards zero as the number of clusters increased.  BC2 appeared to perform slightly better than BC0 for the detailed GEE method.

\section{Discussion}

This paper provides details on a GEE-type method for analyzing correlated binary data and its implementation using the GEECORR SAS macro.  The extended GEE method should be considered when there is interest in the within-cluster correlation structure in addition to the marginal mean regression model.  The additional estimating equation for $\boldsymbol \alpha$ in the extended GEE allows the user to flexibly specify more complex correlation patterns beyond those of GEE but is computationally simpler than GEE2.  Moreover, correlations and their respective variability can be directly estimated using this method.  Three examples demonstrate different study designs when the estimation of correlation parameters may be beneficial.

A version of the more detailed GEE described by \citet{prentice_correlated_1988} has also been incorporated in the GEECORR macro 1.06.  The detailed GEE considers that efficient estimating equations for $\boldsymbol\beta$ and $\boldsymbol\alpha$ are not orthogonal.  The covariance of the estimating equations provides additional information during both parameter estimation and variance estimation.  As shown in our examples and simulation study, both the extended and detailed GEE methods estimated the marginal mean parameters with small to negligible bias and confidence interval coverage near the nominal 95\% level.  Like GEE, marginal mean parameter estimates from the extended and detailed GEE are also robust to misspecification of the working correlation matrix since the estimating equations for $\boldsymbol\beta$ and $\boldsymbol\alpha$ can be separated.

According to our simulation, the primary difference between the extended and detailed GEE methods was the performance of the correlation variance estimators.  BC0 and BC2 from the extended GEE method often overestimated the Monte Carlo simulation correlation variances and the bias increased as the number of clusters increased.  In contrast, BC0 and BC2 from the detailed GEE method underestimated the Monte Carlo simulation correlation variances, but the bias trended towards zero as the number of clusters increased.  
Therefore, when there is specific interest in the correlation parameters, use of the detailed GEE approach should be considered to provide less biased variance estimates, particularly when there are a sufficient number of clusters.

Common issues with software for clustered data also persist with the GEECORR macro.  For instance, the correlation covariate data set may get exponentially larger as the cluster size increases.  Thus, even a study with a small number of large clusters may be computationally intensive due to the matrix operations performed on the correlation covariate data set.  Specifically, in SAS software 9.4, SAS/IML holds all matrices in RAM and has a default maximum matrix memory size of 2GB.  Thus, if computations involve extremely large or too many large matrices, then SAS/IML may not be able to allocate sufficient memory to the processes.  Fortunately, maximum matrix memory size can be increased by changing the SAS software window shortcut or the SAS software config file \citep{wicklin_large_2015}.  Additionally, issues may arise if the covariance matrix for the $\boldsymbol\beta$ estimating equation is not positive definite.  If the covariance matrix is not positive definite, then the GEECORR macro will identify the problematic cluster and exit the call.   

\section{Tables and Figures}

\FloatBarrier

\begin{table}[p]
\begin{center}
	\caption{GEECORR Macro Inputs}\label{table:GEECORROptions}
	\begin{tabular}{ p{2.7cm} p{3.7cm} p{7.3cm} c }
	\hline
Macro Variable & Input* & Description & Required? \\
\hline
XYDATA & Data set & Data set containing the outcome and marginal   mean covariates & Y \\
YVAR & Variable & 0/1 binary outcome variable in xydata & Y \\
XVAR & Variable(s) & Mean covariate variables in xydata & Y \\
ID & Variable & Unique cluster identifier in xydata & Y \\
ZDATA & Data set & Data set containing correlation covariates & Y \\
ZVAR & Variable(s) & Correlation covariate variables in zdata & Y \\
WDATA & Data set & Data set containing cluster weights & Y \\
WVAR & Variable & Frequency weight variable & Y \\
\multirow[t]{3}{*}{LINK} & 1=Identity & \multirow[t]{3}{*}{Link function for the marginal mean model} & \multirow{3}{*}{N} \\
 & 2=Logarithm &  &  \\
 & \textbf{3=Logit} &  &  \\
\multirow[t]{3}{*}{CORRLINK} & \textbf{1=Identity} & \multirow[t]{3}{*}{Link function for the correlation model} & \multirow{3}{*}{N} \\
 & 2=Logarithm &  &  \\
 & 3=Logit &  &  \\
 & 4=Fisher's Z &  &  \\
CLSOUT & Data set & Output data set of cluster deletion diagnostics for beta and alpha & N \\
OBSOUT & Data set & 
\begin{tabular}[c]{@{}l@{}}Output data set of observation deletion\\diagnostics for beta and alpha\end{tabular}
 & N \\
PROBOUT & Data set & Output data set of predicted probabilities of   yvar=1 for each observation & N \\
MAXITER & Number \textbf{(20)} & 
\begin{tabular}[c]{@{}l@{}}Maximum number of iterations for Fisher\\ Scoring\end{tabular}
 & N \\
EPSILON & Number \textbf{(0.00001)} & Tolerance for convergence & N \\
PRINTRANGE & YES, \textbf{NO} & 
\begin{tabular}[c]{@{}l@{}}Print option for additional details of range \\violations from all iterations \end{tabular}
& N \\
SHRINK & THETA, ALPHA & Parameter shrinking option if range violations   occur & N \\
MAKEVONE & YES, \textbf{NO} & If YES, set $\mbox{var}(R_{ijk})=1$ & N \\
\multirow[t]{2}{*}{STARTBETA} & Number(s) & \multirow[t]{2}{*}{Specify initial values for $\boldsymbol\beta$} & \multirow{2}{*}{N} \\
 & \textbf{(Logistic Regression)} &  &  \\
STARTALPHA & Number(s) \textbf{(0.01)} & Specify initial values for $\boldsymbol\alpha$ & N \\
FIXALPHA & YES, \textbf{NO} & If YES, $\boldsymbol\alpha$ is fixed at the value of STARTALPHA & N \\
MOREFITALG & YES, \textbf{NO} & If YES, use the detailed GEE method & N \\
\hline
\multicolumn{4}{l}{*Default inputs are bolded.}
\end{tabular}
\end{center}
\end{table}

\FloatBarrier

\begin{table}[p]
	\begin{center}
		\caption{Mean Parameter Estimates*, Robust Standard Errors (BC0), and Bias-Corrected Standard Errors (BC2) for GTS Study}\label{table:GTSMeanParm}
		\begin{tabular}{lcccccc}
		\hline
		 & \multicolumn{3}{c}{\textit{Logit Link}} & \multicolumn{3}{c}{\textit{Log Link}} \\
		Mean Model & Estimate & BC0 & BC2 & Estimate & BC0 & BC2 \\
		\hline
		Intercept & \textbf{-5.270} & 0.489 & 0.516 & \textbf{-5.252} & 0.482 & 0.509 \\
		Type of Work &  &  &  &  &  &  \\
		\hspace{3mm}Priming & \textbf{1.264} & 0.356 & 0.381 & \textbf{1.222} & 0.348 & 0.372 \\
		\hspace{3mm}Priming/barning & 0.429 & 0.422 & 0.445 & 0.423 & 0.415 & 0.438 \\
		\hspace{3mm}Other or Barning & -0.735 & 0.771 & 0.805 & -0.733 & 0.765 & 0.799 \\
		Less than 5 years of experience & 0.407 & 0.424 & 0.450 & 0.383 & 0.414 & 0.440 \\
		Worked in wet clothes & \textbf{0.930} & 0.384 & 0.403 & \textbf{0.908} & 0.379 & 0.397 \\
		Temperature (1 unit = 10$^{\circ}$F) & \textbf{0.438} & 0.195 & 0.210 & \textbf{0.414} & 0.190 & 0.205 \\
		Tobacco Use & -0.392 & 0.307 & 0.322 & -0.382 & 0.300 & 0.314 \\
		\hline
		\multicolumn{7}{l}{\begin{tabular}[c]{@{}l@{}}*Results are the same for extended and detailed GEE methods to three decimal places.\\
					Bold estimates indicate p-value\textless{}0.05.\end{tabular}}
		\end{tabular}
	\end{center}
\end{table}

\begin{table}[h]
	\begin{center}
		\caption{Estimates and 95\% Confidence Intervals* for GTS Study}\label{table:GTSORRR}
		\begin{tabular}{lcc}
		\hline
		 & Odds Ratios (95\% CI) & Risk Ratios (95\% CI) \\
		 \hline
		Type of Work &  &  \\
		\hspace{3mm}Priming & \textbf{3.54 (1.68, 7.47)} & \textbf{3.39 (1.64, 7.04)} \\
		\hspace{3mm}Priming/barning & 1.54 (0.64, 3.67) & 1.53 (0.65, 3.60) \\
		\hspace{3mm}Other or Barning & 0.48 (0.10, 2.32) & 0.48 (0.10, 2.30) \\
		Less than 5 years of experience & 1.50 (0.62, 3.63) & 1.47 (0.62, 3.47) \\
		Worked in wet clothes & \textbf{2.53 (1.15, 5.58)} & \textbf{2.48 (1.14, 5.40)} \\
		Temperature (1 unit = 10$^{\circ}$F) & \textbf{1.55 (1.03, 2.34)} & \textbf{1.51 (1.01, 2.26)} \\
		Tobacco Use & 0.68 (0.36, 1.27) & 0.68 (0.37, 1.26) \\
		\hline
		\multicolumn{3}{l}{\begin{tabular}[c]{@{}l@{}}*95\% CI computed based on bias-corrected standard errors.\\
					Bold estimates indicate p-value\textless{}0.05.\end{tabular}}
		\end{tabular}
	\end{center}
\end{table}

\begin{table}[h]
	\begin{center}
	\caption{Correlation Parameter Estimates*, Robust Standard Errors (BC0), and Bias-Corrected Standard Errors (BC2) for GTS Study}\label{table:GTSCorrParm}
		\begin{tabular}{l c c c c c c}
		\hline
		 & \multicolumn{3}{c}{Extended GEE} & \multicolumn{3}{c}{Detailed GEE} \\
		Correlation Model & Estimate & BC0 & BC2 & Estimate & BC0 & BC2 \\ 
		\hline
		Within-worker & 0.0261 & 0.0149 & 0.0157 & \textbf{0.0261} & 0.0115 & 0.0120 \\
		Within-camp & 0.0143 & 0.0122 & 0.0136 & 0.0143 & 0.0100 & 0.0113 \\
		\hline
		\multicolumn{7}{l}{\begin{tabular}[c]{@{}l@{}}*Results are the same for mean models with logit and log links to three\\ decimal places.
		Bold estimates indicate p-value\textless{}0.05.\end{tabular}}
		\end{tabular}
	\end{center}
\end{table}

\FloatBarrier

\FloatBarrier

\begin{table}[p]
\begin{center}\caption{Correlation Selection Criteria for CARDIA Study}\label{table:CARDIACorrSelection}
\begin{tabular}{llcccc}
\hline
\multicolumn{1}{c}{\multirow{2}{*}{Race/Sex Group}} & \multicolumn{1}{c}{\multirow{2}{*}{Selection Criterion}} & \multicolumn{4}{c}{Correlation Model} \\ \cline{3-6} 
\multicolumn{1}{c}{} & \multicolumn{1}{c}{} & \multicolumn{1}{c}{Exchangeable} & Modified AR(1) & \multicolumn{1}{c}{M-Dependent} & \multicolumn{1}{c}{Unequal Time} \\
\hline
Black Females & Lg & 5446.384 & 4583.603 & 5547.142 & \textbf{5570.142} \\
\textit{} & CIC & 20.732 & 20.626 & \textbf{20.406} & 20.427 \\
\textit{} & GPC & \textbf{7239.869} & 11604.000 & 7330.455 & 7255.355 \\
\textit{} & TECM & 0.484 & 0.502 & \textbf{0.474} & 0.476 \\
Black Males & Lg & 3608.711 & 3283.590 & \textbf{3787.001} & 3783.174 \\
\textit{} & CIC & 19.709 & 19.697 & \textbf{19.425} & 19.471 \\
\textit{} & GPC & 5341.486 & 7766.588 & 5355.994 & \textbf{5300.181} \\
\textit{} & TECM & 0.576 & 0.605 & 0.570 & \textbf{0.570} \\
White Females & Lg & 5391.115 & 4449.919 & 5478.145 & \textbf{5506.765} \\
\textit{} & CIC & 20.339 & 20.348 & 19.956 & \textbf{19.867} \\
\textit{} & GPC & \textbf{6972.201} & 11366.000 & 7179.408 & 7176.295 \\
\textit{} & TECM & 0.713 & 0.740 & 0.704 & \textbf{0.702} \\
White Males & Lg & 5027.333 & 4680.884 & 5273.232 & \textbf{5282.913} \\
\textit{} & CIC & 21.249 & 21.103 & 20.810 & \textbf{20.779} \\
\textit{} & GPC & \textbf{6402.117} & 9336.887 & 6556.741 & 6469.017 \\
\textit{} & TECM & 0.737 & 0.756 & 0.722 & \textbf{0.720}\\
\hline
\multicolumn{6}{l}{\begin{tabular}[c]{@{}l@{}}*Larger values of Lg and smaller values of CIC, GPC, and TECM suggest better correlation\\ structure.  Bold values indicate the correlation selected by each criteria.\end{tabular}}
\end{tabular}
\end{center}
\end{table}

\FloatBarrier

\FloatBarrier

\begin{table}[h]
	\begin{center}
		\caption{Extended GEE Mean Parameter Estimates and Robust Standard Errors (BC0) for CARDIA Study}\label{table:CARDIAMeanEst}
\begin{tabular}{lcccccccc}
\hline
\textbf{} & \multicolumn{8}{c}{\textit{Exchangeable}} \\  \cline{2-9}
\textbf{} & \multicolumn{2}{c}{\textit{Black Females}} & \multicolumn{2}{c}{\textit{Black Males}} & \multicolumn{2}{c}{\textit{White Females}} & \multicolumn{2}{c}{\textit{White Males}} \\
 & Estimate & BC0 & Estimate & BC0 & Estimate & BC0 & Estimate & BC0 \\
 \hline
Intercept & -0.021 & 0.111 & 0.168 & 0.110 & 0.165 & 0.147 & 0.300 & 0.155 \\
Age/10 & \textbf{0.343} & 0.157 & \textbf{0.654} & 0.177 & -0.245 & 0.187 & -0.099 & 0.197 \\
Age-Square & -0.597 & 0.437 & 0.293 & 0.469 & -0.309 & 0.514 & -0.757 & 0.576 \\
Some College & \textbf{-0.680} & 0.117 & \textbf{-0.949} & 0.125 & \textbf{-0.593} & 0.161 & \textbf{-0.960} & 0.169 \\
College Degree & \textbf{-1.744} & 0.171 & \textbf{-1.795} & 0.177 & \textbf{-1.864} & 0.161 & \textbf{-1.994} & 0.168 \\
Year(Yr/10) & \textbf{0.463} & 0.214 & 0.385 & 0.241 & \textbf{-0.707} & 0.265 & -0.004 & 0.248 \\
Year-square & \textbf{-0.934} & 0.373 & -0.334 & 0.415 & 0.510 & 0.474 & -0.388 & 0.433 \\
Year-cubic & \textbf{0.358} & 0.168 & 0.037 & 0.186 & -0.200 & 0.215 & 0.132 & 0.193 \\
\hline
\textbf{} & \multicolumn{8}{c}{\textit{Modified AR(1)}} \\  \cline{2-9}
\textbf{} & \multicolumn{2}{c}{\textit{Black Females}} & \multicolumn{2}{c}{\textit{Black Males}} & \multicolumn{2}{c}{\textit{White Females}} & \multicolumn{2}{c}{\textit{White Males}} \\
 & Estimate & BC0 & Estimate & BC0 & Estimate & BC0 & Estimate & BC0 \\
 \hline
Intercept & -0.014 & 0.111 & 0.164 & 0.110 & 0.146 & 0.147 & 0.294 & 0.156 \\
Age/10 & \textbf{0.318} & 0.157 & \textbf{0.648} & 0.176 & -0.207 & 0.187 & -0.046 & 0.193 \\
Age-Square & -0.597 & 0.431 & 0.314 & 0.468 & -0.185 & 0.506 & -0.641 & 0.577 \\
Some College & \textbf{-0.681} & 0.117 & \textbf{-0.971} & 0.125 & \textbf{-0.566} & 0.162 & \textbf{-1.006} & 0.169 \\
College Degree & \textbf{-1.775} & 0.171 & \textbf{-1.784} & 0.177 & \textbf{-1.902} & 0.163 & \textbf{-1.986} & 0.167 \\
Year(Yr/10) & \textbf{0.492} & 0.228 & 0.469 & 0.258 & \textbf{-0.695} & 0.280 & -0.037 & 0.258 \\
Year-square & \textbf{-1.009} & 0.392 & -0.467 & 0.439 & 0.496 & 0.496 & -0.286 & 0.447 \\
Year-cubic & \textbf{0.390} & 0.174 & 0.094 & 0.194 & -0.202 & 0.222 & 0.071 & 0.197 \\
\hline
 & \multicolumn{8}{c}{\textit{5-Dependent}} \\   \cline{2-9}
 & \multicolumn{2}{c}{\textit{Black Females}} & \multicolumn{2}{c}{\textit{Black Males}} & \multicolumn{2}{c}{\textit{White Females}} & \multicolumn{2}{c}{\textit{White Males}} \\
\textbf{} & Estimate & BC0 & Estimate & BC0 & Estimate & BC0 & Estimate & BC0 \\
\hline
Intercept & -0.020 & 0.110 & 0.171 & 0.109 & 0.157 & 0.146 & 0.295 & 0.155 \\
Age/10 & \textbf{0.332} & 0.156 & \textbf{0.651} & 0.175 & -0.231 & 0.185 & -0.079 & 0.193 \\
Age-Square & -0.597 & 0.432 & 0.310 & 0.464 & -0.260 & 0.505 & -0.702 & 0.570 \\
Some College & \textbf{-0.679} & 0.117 & \textbf{-0.963} & 0.124 & \textbf{-0.582} & 0.160 & \textbf{-0.981} & 0.168 \\
College Degree & \textbf{-1.758} & 0.169 & \textbf{-1.795} & 0.176 & \textbf{-1.885} & 0.160 & \textbf{-1.995} & 0.166 \\
Year(Yr/10) & \textbf{0.480} & 0.214 & 0.386 & 0.241 & \textbf{-0.704} & 0.266 & -0.014 & 0.244 \\
Year-square & \textbf{-0.963} & 0.371 & -0.327 & 0.416 & 0.512 & 0.475 & -0.348 & 0.429 \\
Year-cubic & \textbf{0.369} & 0.166 & 0.035 & 0.186 & -0.204 & 0.215 & 0.107 & 0.191 \\
\hline
 & \multicolumn{8}{c}{\textit{Unequal Time}} \\   \cline{2-9}
 & \multicolumn{2}{c}{\textit{Black Females}} & \multicolumn{2}{c}{\textit{Black Males}} & \multicolumn{2}{c}{\textit{White Females}} & \multicolumn{2}{c}{\textit{White Males}} \\
\textbf{} & Estimate & BC0 & Estimate & BC0 & Estimate & BC0 & Estimate & BC0 \\
\hline
Intercept & -0.020 & 0.110 & 0.167 & 0.110 & 0.152 & 0.146 & 0.290 & 0.154 \\
Age/10 & \textbf{0.332} & 0.156 & \textbf{0.656} & 0.175 & -0.239 & 0.184 & -0.086 & 0.193 \\
Age-Square & -0.597 & 0.432 & 0.325 & 0.465 & -0.257 & 0.503 & -0.688 & 0.569 \\
Some College & \textbf{-0.679} & 0.117 & \textbf{-0.960} & 0.124 & \textbf{-0.574} & 0.160 & \textbf{-0.980} & 0.168 \\
College Degree & \textbf{-1.758} & 0.169 & \textbf{-1.794} & 0.176 & \textbf{-1.882} & 0.160 & \textbf{-1.984} & 0.166 \\
Year(Yr/10) & \textbf{0.480} & 0.214 & 0.401 & 0.241 & \textbf{-0.701} & 0.266 & -0.012 & 0.245 \\
Year-square & \textbf{-0.963} & 0.371 & -0.356 & 0.414 & 0.508 & 0.475 & -0.353 & 0.428 \\
Year-cubic & \textbf{0.369} & 0.166 & 0.046 & 0.185 & -0.205 & 0.215 & 0.107 & 0.191 \\
\hline
\multicolumn{9}{l}{*Bold estimates indicate p-value\textless{}0.05.}
\end{tabular}
	\end{center}
\end{table}

\FloatBarrier
\begin{table}[]
\begin{center}
\caption{Extended GEE Parameter Estimates and Robust Standard Errors (BC0) for 2013-2014 NHANES, PD$\ge$4mm} \label{table:NHANESResults}
\begin{tabular}{p{7cm}ccc}
\hline
Mean Model & Estimate & BC0 & Odds Ratios* (95\% CI) \\
\hline
Intercept & -3.199 & 0.056 &  -- \\
Maxillary posterior right tooth & 0.016 & 0.046 & 1.02 (0.93, 1.11) \\
Maxillary anterior tooth & \textbf{-0.513} & 0.066 & \textbf{0.60 (0.53, 0.68)} \\
Maxillary posterior left tooth & \textbf{0.183} & 0.046 & \textbf{1.20 (1.10, 1.31)} \\
Mandibular posterior left tooth & -0.080 & 0.044 & 0.92 (0.85, 1.01) \\
Mandibular anterior tooth & \textbf{-0.671} & 0.057 & \textbf{0.51 (0.46, 0.57)} \\
Mesiobuccal site & \textbf{-0.430} & 0.034 & \textbf{0.65 (0.61, 0.69)} \\
Buccal site & \textbf{-1.775} & 0.069 & \textbf{0.17 (0.15, 0.19)} \\
Distal buccal site & \textbf{-0.371} & 0.028 & \textbf{0.69 (0.65, 0.73)} \\
Mesiolingual site & \textbf{-0.209} & 0.018 & \textbf{0.81 (0.78, 0.84)} \\
Lingual site & \textbf{-1.178} & 0.052 & \textbf{0.31 (0.28, 0.34)} \\ \hline
Correlation Model & Estimate & BC0 & \multicolumn{1}{l}{} \\ \hline
Adjacent sites, same tooth or adjacent tooth, and share IP space & \textbf{0.338} & 0.028 &  \\
Adjacent sites, same tooth, and do not share IP space & \textbf{0.295} & 0.029 &  \\
Non-adjacent sites, same tooth & \textbf{0.269} & 0.025 &  \\
Non-adjacent sites, adjacent teeth & \textbf{0.234} & 0.023 &  \\
Sites on vertically adjacent teeth & \textbf{0.178} & 0.021 &  \\
Sites on non-adjacent teeth & \textbf{0.148} & 0.019 &  \\ \hline
\multicolumn{4}{l}{\begin{tabular}[c]{@{}l@{}}*Mandibular posterior right tooth and distal lingual sites are the reference groups\\ for tooth sextant and tooth site, respectively. Bold estimates indicate p-value\textless{}0.05.\end{tabular}}
\end{tabular}
\end{center}
\end{table}

\FloatBarrier

\begin{table}[]
\begin{center}
\caption{Simulated estimates, coverage of robust (BC0) and bias-corrected (BC2) variances, and percent relative bias of $\boldsymbol\beta$ and $\boldsymbol\alpha$ from 1000 simulations of $K$ clusters 
[$\beta = (-3.9, 1.3, 0.4, -0.7, 0.4, 0.9, 0.4, -0.4)'$ and $\alpha=(0.03, 0.01)'$]} \label{table:SimResults}
\begin{tabular}{cccccccc}
\hline
 \multicolumn{8}{c}{\textit{Marginal Mean Parameters, Either method}} \\ \hline
\multicolumn{2}{c}{\textit{}} & \multicolumn{2}{c}{\textit{Estimate}} & \multicolumn{2}{c}{\textit{\begin{tabular}[c]{@{}c@{}}BC0\end{tabular}}} & \multicolumn{2}{c}{\textit{\begin{tabular}[c]{@{}c@{}}BC2\end{tabular}}} \\ 
$K$ & Parameter & Estimate & \% Bias & Coverage & \% Bias & Coverage & \% Bias \\ \hline
\multicolumn{1}{r}{37} & $\beta_0$ & -3.937 & 0.95 & 92.22 & -11.93 & 93.55 & -1.64 \\
 & $\beta_1$ & 1.302 & 0.15 & 92.12 & -9.00 & 94.06 & 4.41 \\
 & $\beta_2$ & 0.393 & -1.86 & 92.32 & -6.23 & 94.47 & 5.64 \\
 & $\beta_3$ & -0.733 & 4.75 & 93.76 & -6.98 & 94.88 & 2.15 \\
 & $\beta_4$ & 0.413 & 3.34 & 92.53 & -14.63 & 94.06 & -0.71 \\
 & $\beta_5$ & 0.906 & 0.69 & 93.96 & -7.40 & 95.09 & 2.40 \\
 & $\beta_6$ & 0.412 & 2.90 & 93.24 & -10.87 & 94.17 & -1.30 \\
 & $\beta_7$ & -0.400 & -0.09 & 92.22 & -16.70 & 93.55 & -5.77 \\
\multicolumn{1}{r}{74} & $\beta_0$ & -3.937 & 0.94 & 92.90 & -10.46 & 93.20 & -5.50 \\
 & $\beta_1$ & 1.309 & 0.69 & 94.20 & -4.23 & 94.70 & 2.40 \\
 & $\beta_2$ & 0.404 & 0.99 & 94.70 & 0.62 & 95.20 & 6.67 \\
 & $\beta_3$ & -0.720 & 2.83 & 94.50 & -0.73 & 95.20 & 3.94 \\
 & $\beta_4$ & 0.412 & 2.94 & 93.60 & -6.50 & 94.60 & 0.56 \\
 & $\beta_5$ & 0.916 & 1.77 & 93.70 & -10.94 & 94.30 & -6.41 \\
 & $\beta_6$ & 0.407 & 1.65 & 92.50 & -11.07 & 93.30 & -6.55 \\
 & $\beta_7$ & -0.408 & 2.03 & 93.60 & -3.53 & 94.10 & 2.38 \\
111 & $\beta_0$ & -3.919 & 0.48 & 95.30 & 1.31 & 95.30 & 4.98 \\
 & $\beta_1$ & 1.308 & 0.64 & 95.00 & 1.31 & 95.40 & 5.87 \\
 & $\beta_2$ & 0.404 & 1.11 & 94.80 & -0.03 & 94.90 & 3.91 \\
 & $\beta_3$ & -0.709 & 1.28 & 94.50 & 0.49 & 94.80 & 3.61 \\
 & $\beta_4$ & 0.402 & 0.46 & 95.50 & 2.83 & 95.90 & 7.86 \\
 & $\beta_5$ & 0.909 & 1.01 & 94.10 & 0.05 & 94.40 & 3.40 \\
 & $\beta_6$ & 0.400 & 0.09 & 94.40 & -5.81 & 94.90 & -2.66 \\
 & $\beta_7$ & -0.398 & -0.47 & 94.20 & -2.20 & 94.90 & 1.70 \\ \hline 
 \multicolumn{8}{c}{\textit{Correlation Parameters, Extended GEE}} \\ \hline
\multicolumn{2}{c}{\textit{}} & \multicolumn{2}{c}{\textit{Estimate}} & \multicolumn{2}{c}{\textit{\begin{tabular}[c]{@{}c@{}}BC0\end{tabular}}} & \multicolumn{2}{c}{\textit{\begin{tabular}[c]{@{}c@{}}BC2\end{tabular}}} \\
$K$ & Parameter & Estimate & \% Bias & Coverage & \% Bias & Coverage & \% Bias \\ \hline
37 & $\alpha_1$ & 0.027 & -10.20 & 79.53 & 2.71 & 79.94 & 10.60 \\
 & $\alpha_2$ & 0.009 & -13.17 & 78.40 & -8.42 & 79.32 & 1.77 \\
74 & $\alpha_1$ & 0.028 & -5.12 & 86.50 & 16.09 & 86.80 & 20.63 \\
 & $\alpha_2$ & 0.009 & -8.62 & 80.90 & -0.89 & 81.40 & 4.42 \\
111 & $\alpha_1$ & 0.030 & -1.63 & 90.60 & 18.94 & 90.80 & 22.07 \\
 & $\alpha_2$ & 0.010 & -4.91 & 86.10 & 8.29 & 86.60 & 12.33 \\ \hline 
\multicolumn{8}{c}{\textit{Correlation Parameters, Detailed GEE}} \\ \hline
\multicolumn{2}{c}{\textit{}} & \multicolumn{2}{c}{\textit{Estimate}} & \multicolumn{2}{c}{\textit{\begin{tabular}[c]{@{}c@{}}BC0\end{tabular}}} & \multicolumn{2}{c}{\textit{\begin{tabular}[c]{@{}c@{}}BC2\end{tabular}}} \\ 
$K$ & Parameter & Estimate & \% Bias & Coverage & \% Bias & Coverage & \% Bias \\ \hline
37 & $\alpha_1$ & 0.027 & -10.20 & 77.69 & -24.87 & 78.40 & -18.98 \\
 & $\alpha_2$ & 0.009 & -13.17 & 78.20 & -24.27 & 79.22 & -15.49 \\
74 & $\alpha_1$ & 0.028 & -5.12 & 84.70 & -11.40 & 85.10 & -7.91 \\
 & $\alpha_2$ & 0.009 & -8.62 & 80.50 & -15.02 & 80.90 & -10.33 \\
111 & $\alpha_1$ & 0.030 & -1.63 & 88.50 & -8.27 & 88.70 & -5.83 \\
 & $\alpha_2$ & 0.010 & -4.91 & 85.60 & -5.52 & 86.10 & -1.90 \\ \hline
\end{tabular}
\end{center}
\end{table}

\FloatBarrier

\appendix
\section{Derivation of Joint Distribution}\label{appendix:jointdistderivation}
Modified from \citet{prentice_correlated_1988} Appendix 1. 
Under regularity condition and using a Taylor expansion, the joint distribution of 
$K^{1/2}(\hat{\beta} - \beta)$, $K^{1/2}(\hat{\alpha} - \alpha)$ could be approximated by
\begin{equation} \label{alphabetajointdist}
	\begin{Bmatrix}
	K^{1/2}(\hat{\beta} - \beta)\\
	K^{1/2}(\hat{\alpha} - \alpha)
	\end{Bmatrix}
	\approx
	\begin{bmatrix}
	-K^{-1}\partial U_\beta/\partial\beta & -K^{-1}\partial U_\beta/\partial\alpha\\
	-K^{-1}\partial U_\alpha/\partial\beta  & -K^{-1}\partial U_\alpha/\partial\alpha\\
	\end{bmatrix}^{-1}
	\begin{bmatrix}
	K^{-1/2}U_\beta\\
	K^{-1/2}U_\alpha
	\end{bmatrix}
\end{equation}

Under mild regularity, as $K \rightarrow \infty$, the information matrix can be written as
\begin{equation}\label{jointinformation}
	\begin{bmatrix}
	-K^{-1}\partial U_\beta/\partial\beta & -K^{-1}\partial U_\beta/\partial\alpha\\
	-K^{-1}\partial U_\alpha/\partial\beta  & -K^{-1}\partial U_\alpha/\partial\alpha\\
	\end{bmatrix}
	=
	\begin{bmatrix}
		K^{-1} \sum\limits_{i=1}^{K} D_i'  V_i^{-1} D_i + o_p(1) & o_p(1) \\
		-K^{-1} \sum\limits_{i=1}^{K} E_i'  W_i^{-1} \frac{\partial R_i}{\partial \beta}  + o_p(1) & K^{-1} \sum\limits_{i=1}^{K} E_i'  W_i^{-1} E_i + o_p(1)\\
	\end{bmatrix}
\end{equation}

Therefore \eqref{alphabetajointdist} can be approximated using \eqref{jointinformation} as,
\begin{equation}
	\begin{Bmatrix}
	(\hat{\beta} - \beta)\\
	(\hat{\alpha} - \alpha)
	\end{Bmatrix}
	\approx
	\begin{bmatrix}
	\sum\limits_{i=1}^{K} D_i'  V_i^{-1} D_i  & 0 \\
	-\sum\limits_{i=1}^{K} E_i'  W_i^{-1} \frac{\partial R_i}{\partial \beta}  & \sum\limits_{i=1}^{K} E_i'  W_i^{-1} E_i\\
	\end{bmatrix}^{-1}
	\begin{bmatrix}
	U_\beta\\
	U_\alpha
	\end{bmatrix}
\end{equation}

This implies the fitting algorithm in Section 2.4 can be re-written using the joint distribution approximation as
\begin{equation} \label{appendixfittingalg}
	\begin{bmatrix}
	\hat{\beta}_{s+1}\\
	\hat{\alpha}_{s+1}
	\end{bmatrix}
	=
	\begin{bmatrix}
	\hat{\beta}_{s}\\
	\hat{\alpha}_{s}
	\end{bmatrix}
	+
	\begin{bmatrix}
	\sum\limits_{i=1}^{K} D_i'  V_i^{-1} D_i  & 0 \\
	-\sum\limits_{i=1}^{K} E_i'  W_i^{-1} \frac{\partial R_i}{\partial \beta}  & \sum\limits_{i=1}^{K} E_i'  W_i^{-1} E_i\\
	\end{bmatrix}^{-1}
	\begin{bmatrix}
	U_\beta\\
	U_\alpha
	\end{bmatrix}
\end{equation}

To obtain the variance estimator of the joint asymptotic distribution, consider the linear functions $[K^{1/2} U_\beta, K^{1/2} U_\alpha]$ from \eqref{alphabetajointdist} are asymptotically normally distributed
as $K \rightarrow \infty$, with mean 0 and variance
\begin{equation}\label{jointscorevariance}
	\lim_{i \to \infty} K^{-1}
	\begin{bmatrix}
	D_i' V_i^{-1}cov(Y_i)V_i^{-1}D_i & D_i' V_i^{-1}cov(Y_i,R_i)W_i^{-1}E_i\\
	E_i' W_i^{-1}cov(R_i,Y_i)V_i^{-1}D_i  & E_i' W_i^{-1}cov(R_i)W_i^{-1}E_i\\
	\end{bmatrix}
\end{equation}

Additionally, the matrix inverse of \eqref{jointinformation} converges in probability as $K \rightarrow \infty$ to
\begin{equation}\label{jointinformationinverse}
	\begin{bmatrix}
	A & 0\\
	B  & C\\
	\end{bmatrix}
\end{equation}
where
$$
A = \left(\sum_{i=1}^K D_i' V_i^{-1} D_i \right)^{-1} 
$$
$$ 
B =\left(\sum_{i=1}^K E_i' W_i^{-1} E_i \right)^{-1}
	\left(\sum_{i=1}^K E_i' W_i^{-1} \partial R_i/\partial\beta \right) 
	\left(\sum_{i=1}^K D_i' V_i^{-1} D_i\right)^{-1}
$$
$$
C =  \left(\sum_{i=1}^K E_i' W_i^{-1} E_i \right)^{-1}
$$

Combining \eqref{jointscorevariance} and \eqref{jointinformationinverse}, and evaluating \eqref{jointscorevariance} at consistent estimates for $(\beta,\alpha)$ denoted by $(\hat{\beta}, \hat{\alpha})$, the variance of the joint distribution of $K^{1/2}(\hat{\beta} - \beta)$, $K^{1/2}(\hat{\alpha} - \alpha)$  is consistently estimated by

\begin{equation}
	\begin{bmatrix}
	A & 0\\
	B  & C\\
	\end{bmatrix}
	\begin{bmatrix}
	\Lambda_{11} & \Lambda_{12}\\
	\Lambda_{21}  & \Lambda_{22}\\
	\end{bmatrix}
	\begin{bmatrix}
	A & B'\\
	0  & C\\
	\end{bmatrix}
\end{equation}
where
$$
\Lambda_{11} = \sum\limits_{i=1}^{K} D_i' V_i^{-1}(Y_i - \mu_i)(Y_i - \mu_i)'V_i^{-1}D_i 
$$
$$ 
\Lambda_{12} =  \sum\limits_{i=1}^{K} D_i' V_i^{-1}(Y_i-\mu_i)(R_i-\rho_i)'W_i^{-1}E_i
$$
$$
\Lambda_{22} = \sum\limits_{i=1}^{K} E_i' W_i^{-1}(R_i-\rho_i)(R_i-\rho_i)'W_i^{-1}E_i
$$
$$
\Lambda_{21}=\Lambda_{12}'
$$

Note that,
\begin{align*}
\frac{\partial R_i}{\partial \beta} =
				-\{
				D_{ij}(Y_{ij} - \mu_{ij}) + D_{ik}(Y_{ik} - \mu_{ik})
				- \frac{1}{2} (Y_{ij} - \mu_{ij}) (Y_{ik} - \mu_{ik}) 
				[(1 - 2\mu_{ij}) D_{ij} \mu_{ij}^{-1} (1 - \mu_{ij})^{-1} \\
					+ (1 - 2\mu_{ik}) D_{ik} \mu_{ik}^{-1} (1 - \mu_{ik})^{-1}]					
				\} /
				\{\mu_{ij}(1 - \mu_{ij})\mu_{ik}(1 - \mu_{ik})\}^{1/2}
\end{align*}

Since $\partial R_i / \partial \beta$ is a part of the variance of the score equation, it is reasonable to use the expected information rather than the observed information in \eqref{appendixfittingalg} and \eqref{jointinformationinverse}, where

$$
E \left[ \frac{\partial R_i}{\partial \beta} \right] =
				-\{
				D_{ij}(Y_{ij} - \mu_{ij}) + D_{ik}(Y_{ik} - \mu_{ik})					
				\} /
				\{\mu_{ij}(1 - \mu_{ij})\mu_{ik}(1 - \mu_{ik})\}^{1/2}
$$

If $B=0$, that is, the estimating equations of $\boldsymbol\beta$ and $\boldsymbol\alpha$ are independent, then the fitting algorithm reduces to the usual Prentice (1988) extended GEE approach described in section 2.3.

\section{Sample Programming}

The software is available at \url{http://www.bios.unc.edu/~preisser/personal/software.html}.  Sample code to develop the three input data sets is provided for the GTS data.

GEECORR macro invocations from all examples are also provided.

\subsection{GTS Data: Partial Listing and Data Preprocessing}\label{appendix:dataprocessingcode}

\noindent Partial listing of GTS data:
\begin{table}[h]
\tiny
\begin{tabular}{rrrrrrrrrrrrr}
\hline
\textbf{Obs} & \textbf{SITEID} & \textbf{NEWPID} &\textbf{unitid} & \textbf{GTS} & \textbf{one} & \textbf{workty1} & \textbf{workty2} & \textbf{workty4} & \textbf{experien\_lt5} & \textbf{wet} & \textbf{temp92\_10} & \textbf{tobuser} \\ \hline
\textbf{31} & 1008 & 158 & 31 & 0 & 1 & 0 & 1 & 0 & 0 & 0 & -0.1 & 1 \\
\textbf{32} & 1008 & 158 & 32 & 0 & 1 & 0 & 1 & 0 & 0 & 1 & -0.5 & 1 \\
\textbf{33} & 1008 & 158 & 33 & 0 & 1 & 0 & 1 & 0 & 0 & 1 & -1.2 & 1 \\
\textbf{34} & 1008 & 158 & 34 & 0 & 1 & 0 & 1 & 0 & 0 & 1 & -1.1 & 1 \\
\textbf{35} & 1008 & 158 & 35 & 0 & 1 & 0 & 1 & 0 & 0 & 0 & -0.8 & 1 \\
\textbf{36} & 1008 & 159 & 36 & 0 & 1 & 0 & 0 & 0 & 0 & 1 & -0.4 & 1 \\
\textbf{37} & 1008 & 159 & 37 & 0 & 1 & 0 & 0 & 0 & 0 & 1 & 0.0 & 1 \\
\textbf{38} & 1008 & 159 & 38 & 0 & 1 & 0 & 0 & 0 & 0 & 1 & 0.3 & 1 \\
\textbf{39} & 1008 & 159 & 39 & 0 & 1 & 0 & 0 & 0 & 0 & 1 & 0.8 & 1 \\
\textbf{40} & 1008 & 159 & 40 & 0 & 1 & 0 & 0 & 0 & 0 & 1 & 0.8 & 1 \\
\hline
\end{tabular}
\end{table}

\noindent Code to produce correlation covariate data set:
\begin{verbatim}
proc sql;
     create table zd (drop=i j k) as
     select
          a1.siteid as i,
          a1.unitid as j,
          a2.unitid as k,
          (a1.newpid = a2.newpid) as worker,     /*same worker*/
          (a1.newpid ne a2.newpid) as camp    /*different worker*/
     from 
          xy as a1,
          xy as a2
     where 
          a1.unitid < a2.unitid and a1.siteid eq a2.siteid
     order by i, j, k;
quit;
\end{verbatim}

\noindent Partial listing of correlation covariate data set:
\begin{table}[h]
\tiny
\begin{tabular}{rrr}
\hline
\textbf{Obs} & \textbf{WORKER} & \textbf{CAMP} \\ \hline
\textbf{31} & 1 & 0 \\
\textbf{32} & 1 & 0 \\
\textbf{33} & 1 & 0 \\
\textbf{34} & 1 & 0 \\
\textbf{35} & 0 & 1 \\
\textbf{36} & 0 & 1 \\
\textbf{37} & 0 & 1 \\
\textbf{38} & 0 & 1 \\
\textbf{39} & 0 & 1 \\
\textbf{40} & 0 & 1 \\ \hline
\end{tabular}
\end{table}

\noindent Code to create weight data set:
\begin{verbatim}
proc sql;
     create table w as
     select distinct(siteid), 1 as one
     from xy
     order by siteid;
quit;
\end{verbatim}

\noindent Partial listing of weight data set:
\begin{table}[h]
\tiny
\begin{tabular}{rrr}
\hline
\textbf{Obs} & \textbf{SITEID} & \textbf{one} \\ \hline
\textbf{1} & 1008 & 1 \\
\textbf{2} & 1009 & 1 \\
\textbf{3} & 1010 & 1 \\
\textbf{4} & 1012 & 1 \\
\textbf{5} & 1015 & 1 \\ \hline
\end{tabular}
\end{table}

\subsection{GTS Data: Macro Invocation for Mean Model with Logit Link, Extended GEE}\label{appendix:GTSLogitCode}
\begin{verbatim}
%geecorr(xydata = xy,
          yvar = gts,
          xvar = one workty1 workty2 workty4 experien_lt5 wet temp92_10 tobuser,
          id = siteid,
          zdata = zd,
          zvar = worker camp,
          wdata = w,
          wvar = one,
          CLSOUT = gts_cluster_diag,
          PROBOUT = gts_pred,
          printrange = YES);
          quit;
\end{verbatim}

\subsection{GTS Data: Macro Invocation for Mean Model with Logit Link, Detailed GEE}\label{appendix:GTSLogitDetailedCode}
\begin{verbatim}
%geecorr(xydata = xy,
          yvar = gts,
          xvar = one workty1 workty2 workty4 experien_lt5 wet temp92_10 tobuser,
          id = siteid,
          zdata = zd,
          zvar = worker camp,
          wdata = w,
          wvar = one,
          printrange = YES,
          morefitalg = YES);
          quit;
\end{verbatim}

\subsection{GTS Data: Macro Invocation for Mean Model with Log Link Fitted, Extended GEE}\label{appendix:GTSLogCode}
\begin{verbatim}
%geecorr(xydata = xy,
          yvar = gts,
          xvar = one workty1 workty2 workty4 experien_lt5 wet temp92_10 tobuser,
          id = siteid,
          zdata = zd,
          zvar = worker camp,
          wdata = w,
          wvar = one,
          printrange = YES,
          link = 2);
          quit;
\end{verbatim}

\subsection{GTS Data: Macro Invocation for Mean Model with Log Link Fitted, Detailed GEE}\label{appendix:GTSLogDetailedCode}
\begin{verbatim}
%geecorr(xydata = xy,
          yvar = gts,
          xvar = one workty1 workty2 workty4 experien_lt5 wet temp92_10 tobuser,
          id = siteid,
          zdata = zd,
          zvar = worker camp,
          wdata = w,
          wvar = one,
          printrange = YES,
          link = 2,
          morefitalg = YES);
          quit;
\end{verbatim}

\subsection{CARDIA Data: Macro Invocation for Exchangeable Correlation Model, Extended GEE}\label{appendix:CARDIAExchCode}
\begin{verbatim}
%geecorr(xydata = xy_bm,
          yvar = smoke,
          xvar = int age10 age2 smcl c_degree time time2 time3,
          id = pid,
          zdata = z_bm,
          zvar = one,
          wdata = w_bm,
          wvar = one,
          printrange = YES);
          quit;
\end{verbatim}

\subsection{CARDIA Data: Macro Invocation for AR(1) Correlation Model, Extended GEE}\label{appendix:CARDIAARCode}
\begin{verbatim}
%geecorr(xydata = xy_bm,
          yvar = smoke,
          xvar = int age10 age2 smcl c_degree time time2 time3,
          id = pid,
          zdata = z_bm,
          zvar = ztime,
          wdata = w_bm,
          wvar = one,
          printrange = YES,
          corrlink = 2,
          startalpha = -0.2);
          quit;
\end{verbatim}

\subsection{CARDIA Data: Macro Invocation for M-Dependent Correlation Model, Extended GEE}\label{appendix:CARDIAMDepCode}
\begin{verbatim}
%geecorr(xydata = xy_bm,
          yvar = smoke,
          xvar = int age10 age2 smcl c_degree time time2 time3,
          id = pid,
          zdata = z1 z2 z3 z4 z5,
          zvar = ztime,
          wdata = w_bm,
          wvar = one,
          printrange = YES);
          quit;
\end{verbatim}

\subsection{CARDIA Data: Macro Invocation for Unequal Time Correlation Model, Extended GEE}\label{appendix:CARDIAUneqCode}
\begin{verbatim}
%geecorr(xydata = xy_bm,
          yvar = smoke,
          xvar = int age10 age2 smcl c_degree time time2 time3,
          id = pid,
          zdata = z2 z3 z5 z7 z8 z10 z13 z15,
          zvar = ztime,
          wdata = w_bm,
          wvar = one,
          printrange = YES);
          quit;
\end{verbatim}

\subsection{NHANES Data: Macro Invocation for Extended GEE}\label{appendix:NHANESCode}
\begin{verbatim}
%geecorr(xydata = xy_d4_pd4,
          yvar = pd4,
          xvar = one MAX_POSTERIOR_R MAX_ANTERIOR MAX_POSTERIOR_L
                    MAND_POSTERIOR_L MAND_ANTERIOR MF BF DF ML BL,
          id = seqn,
          zdata = corr55,
          zvar = z1316 z14 z15 z17 z1819 z20,
          wdata = nhanes2013_anl_weights,
          wvar = WT_SCALED,
          printrange=YES);
          quit;
\end{verbatim}


\bibliography{geecorr_ref}

\end{document}